\documentclass[aps,showpacs,superscriptaddress,axodraw]{revtex4-2}  
\usepackage{graphicx,multirow,subfig}
\usepackage{float}
\usepackage{bm}
\usepackage{braket}
\usepackage{amsmath}
\usepackage{amssymb}
\usepackage{amscd}
\usepackage{latexsym}
\usepackage{slashed}
\usepackage{color}
\usepackage{graphicx}
\usepackage[normalem]{ulem}
\usepackage{color}
\definecolor{orange}{cmyk}{0,0.5,1,0}
\usepackage{verbatim}
\usepackage{rotating}
\usepackage{diagbox}
\usepackage{cancel}
\usepackage[utf8]{inputenc}
\usepackage{array}
\usepackage{caption}
\usepackage{axodraw}
\usepackage{hyperref}

\def\lsim{\raise0.3ex\hbox{$\;<$\kern-0.75em\raise-1.1ex\hbox{$\sim\;$}}}
\def\gsim{\raise0.3ex\hbox{$\;>$\kern-0.75em\raise-1.1ex\hbox{$\sim\;$}}}

\newcolumntype{L}[1]{>{\raggedright\let\newline\\\arraybackslash\hspace{0pt}}m{#1}}
\newcolumntype{C}[1]{>{\centering\let\newline\\\arraybackslash\hspace{0pt}}m{#1}}
\newcolumntype{R}[1]{>{\raggedleft\let\newline\\\arraybackslash\hspace{0pt}}m{#1}}

\def\be{\begin{equation}}
\def\ee{\end{equation}}
\def\bea{\begin{eqnarray}}
\def\eea{\end{eqnarray}}

\newcommand{\xdownarrow}[1]{%
	{\left\downarrow\vbox to #1{}\right.\kern-\nulldelimiterspace}
}

\begin{document}
\title{Multi-component Dark Matter in a Simplified E$_6$SSM Model}

\author{Shaaban Khalil}
\email[]{skhalil@zewailcity.edu.eg}
\affiliation{Center for Fundamental Physics, Zewail City of Science and Technology, 6 October City, Giza 12588, Egypt}

\author{Stefano Moretti}
\email[]{s.moretti@soton.ac.uk}
\affiliation{\small School of Physics and Astronomy, University of Southampton,
	Southampton, SO17 1BJ, United Kingdom}
	
\author{Diana Rojas-Ciofalo}
\email[]{D.Rojas-Ciofalo@soton.ac.uk}
\affiliation{\small School of Physics and Astronomy, University of Southampton,
	Southampton, SO17 1BJ, United Kingdom}

\author{Harri Waltari.}
\email[]{h.waltari@soton.ac.uk}
\affiliation{\small School of Physics and Astronomy, University of Southampton,
	Southampton, SO17 1BJ, United Kingdom}
\affiliation{\small Particle Physics Department, Rutherford Appleton Laboratory, Chilton, Didcot, Oxon OX11 0QX, UK}


\begin{abstract}
We study Dark Matter (DM) in the Exceptional Supersymmetric Standard Model (E$_6$SSM). The model has both active and inert Higgs superfields and by imposing discrete symmetries one can generate two DM candidates. We show that the lightest higgsinos of the active and inert sectors give a viable setup for two-component DM. We also illustrate the scope of both direct and indirect detection experiments in extracting such a DM sector. Future experiments of the former kind have a good chance of finding the active component while the inert higgsino will be very hard to detect while those of the latter kind will have no sensitivity to either candidate.
	\end{abstract}
\maketitle

\section{Introduction}

Recent Planck satellite observations of the fluctuations in the Cosmic Microwave Background (CMB) \cite{Ade2016,Aghanim2018} confirmed that the largest part of our universe consists of invisible matter: Dark Matter (DM) for $26.8\%$ and Dark Energy (DE) for $68.3\%$ of it, while less than 
$5\%$ of it is in the form of observable matter. Such a small part of visible matter is composed of (anti)quarks and (anti)leptons, in addition to gauge bosons. Therefore, it is not unrealistic to imagine that the DM sector is not minimal either and the assumption of multi-component DM is quite justified. 

In multi-component DM scenarios, we may have a combination of cold and warm DM that could explain the problem of small scale structure, where a discrepancy between collisionless cold DM and observational data was found \cite{Wang2017}. Moreover, having multiple DM particles may provide interesting solutions for avoiding stringent constraints imposed nowadays from negative searches for DM at Direct Detection (DD) and Indirect Detection (ID) experiments and also at the Large Hadron Collider (LHC). 

A DM candidate is natural in the context of Supersymmetry (SUSY) with so-called $R$-parity conservation \cite{ShaabanKhalil2019}. However, the minimal version of SUSY, the so-called 
 Minimal Supersymmetric Standard Model (MSSM), contains only one DM candidate that has been widely studied in the literature. However, the combined LHC and relic abundance constraints
rule out most of the MSSM parameter space except very narrow regions. Therefore, non-minimal SUSY models with a richer structure than the MSSM and hallmark signatures, such as the Exceptional Supersymmetry Standard Model  (E$_6$SSM) of refs. \cite{King2006,King2006a,King2020},  including in its constrained version \cite{Athron2009,Athron2009a,Athron2011,Athron2012,Athron2016,Athron2017},  may provide new DM candidates that account for the observed  relic density without a conflict with other experimental constraints. The E$_6$SSM is a string inspired SUSY scenario with an $E_6$ gauge group. The $E_6$  gauge symmetry prevailing at the scale of a Grand Unification Theory (GUT) 
is then broken via $E_6 \to SU(3)_c\times SU(2)_L\times U(1)_Y\times U(1)_N$ at lower energies. At such scales,  wherein the E$_6$SSM  is essentially a  Standard Model
(SM) $\times$ $U(1)_N$ effective description from the viewpoint of a gauge theory, extra Right-Handed (RH) neutrinos are uncharged under $U(1)_N$ and can then acquire large intermediate scale Majorana masses leading to a Type-I see-saw mechanism to explain the small Left-Handed (LH) neutrino masses \cite{Hall2011}.  

In the E$_6$SSM, the  $4\times 4$  neutralino mass matrix of the MSSM, composed of the bino, the neutral wino  and two active higgsinos, is greatly enlarged into a 12$\times$12 matrix, which further includes 4 inert higgsinos, one active singlino, two inert singlinos and another bino, wherein inert refers to
(the SUSY counterpart of) a (pseudo)scalar field which does not acquire a Vacuum Expectation Value (VEV), unlike an active one which does (i.e., it is a Higgs state). It has been observed that the 6 inert states tend to decouple from the rest of the neutralino spectrum and it makes sense to consider their 6$\times$6 matrix separately. One of the possible scenarios interesting to study then is two-component DM, with one active neutralino and one inert neutralino being the two DM candidates, which is indeed the aim of this study.

The plan of this paper is as follows. In the next section, we describe the simplified version of the E$_6$MSSM we will be dealing with, with two subsections specifically dedicated to describe both active and inert neutralino and neutral (pseudo)scalar fields. Then, in Sect.~\ref{sec:DM}, we discuss the ensuing DM sector and the relic densities. We present our results for  DD and ID rates  in Sects. \ref{sec:DD} and \ref{sec:ID}. We conclude in Sect.~\ref{sec:summa}.

\section{Simplified E$_6$SSM Model}
\label{sec:model}
A Supersymmetric $E_6$ GUT model emerges  from ten-dimensional heterotic string theory after the compactification of extra dimensions.  The $E_6$ gauge group can be broken down to the SM gauge group as follows:
\begin{eqnarray}
E_6 ~ &\longrightarrow& ~ SO(10) \times U(1)_\psi \nonumber\\
        ~ &\longrightarrow& ~ SU(5) ~ \times U(1)_\chi \times U(1)_\psi\nonumber\\
        ~ &\longrightarrow& ~ SU(3)_C \times SU(2)_L \times U(1)_Y \times U(1)_\chi \times U(1)_\psi .
\end{eqnarray}
The low energy gauge group obtained is thus a scenario with the SM gauge group extended by an additional $U(1)_N$ symmetry. This $U(1)_N$
structure  is given by 
\begin{equation} 
U(1)_N = \cos\vartheta\; U(1)_\chi + \sin\vartheta\; U(1)_{\psi},
\end{equation}
where $\tan\vartheta = \sqrt{15}$ so the RH neutrinos are chargeless.
In this case, the fundamental representation of $E_6$, $27_i$-plet, $i=1,2,3$, is decomposed under $SU(5) \times U(1)_N$ as 
\be 
27_i \to (10, \frac{1}{\sqrt{40}})_i + (\bar{5}, \frac{2}{\sqrt{40}})_i + (\bar{5}, \frac{-3}{\sqrt{40}})_i  + (5, \frac{-2}{\sqrt{40}})_i + (1, \frac{5}{\sqrt{40}})_i + (1,0)_i ,
\ee
where the following field associations can be made:
\begin{itemize}
 \item $(10, \frac{1}{\sqrt{40}})_i$ and $(\bar{5}, \frac{2}{\sqrt{40}})_i$ $\rightarrow$ Normal matter
 \item $(\bar{5}, \frac{-3}{\sqrt{40}})_i$ and $(5, \frac{-2}{\sqrt{40}})_i$ $\rightarrow$  Three generations of Higgs doublets $H_{di}, H_{ui}$ and exotic coloured states $\bar{D}_i, D_i$
 \item $(1, \frac{5}{\sqrt{40}})_i$ $\rightarrow$  Three generations of singlets $S_i$
 \item $(1,0)_i$ $\rightarrow$  RH neutrinos
\end{itemize}

At low energies, the $U(1)_N$ is broken spontaneously by the singlet, $S_3$, which develops a VEV, $s$, radiatively. Therefore, we have a $Z^{\prime}$ boson of mass of order of the SUSY breaking scale, say, a few TeV. Automatic anomaly cancellation is ensured by allowing three complete 27 representations of $E_6$ to survive down to the low energy scale. These three 27 representations contain not only the three matter generations but also the Higgs doublets and singlet that will acquire VEVs. Thus, we have other two copies of doublet
 and singlet fields, $H_{d\alpha}, H_{u\alpha}, S_\alpha, \alpha= 1,2$, that do not develop a VEV and hence are inert (or dark) (pseudo)scalars. Their Yukawa couplings to SM matter are consequently very suppressed and this prevents Flavour Changing Neutral Currents
(FCNCs). In this regards, the following VEVs are acquired by the third generation of fields  in the Higgs sector:
 \be 
\langle H^0_{d3}\rangle = \frac{v_d}{\sqrt{2}} = \frac{v\cos\beta}{\sqrt{2}}, \qquad \langle H^0_{u3}\rangle = \frac{v_u}{\sqrt{2}} = \frac{v\sin\beta}{\sqrt{2}}, \qquad \langle S_3\rangle = \frac{s}{\sqrt{2}}.
\ee
In addition to the gauge symmetries, the following discrete symmetries are assumed in this class of models \cite{Hall2011,Hall2009,Athron2016,Athron2017} (see Tab.~\ref{symmetries}).
\begin{itemize}
 \item $Z^H_2$: to distinguish between the third active generation and the inert generations of doublets and singlets, which  supresses flavour transitions. (Note that this symmetry also suppresses $\lambda_{ijk}$ couplings of the forms $\lambda_{\alpha 33}, \lambda_{3\alpha 3}$ and $\lambda_{\alpha\beta\gamma}$ with $\alpha, \beta, \gamma = {1,2}$.)  
\item $Z^L_2$ or $Z^B_2$: to forbid proton decay, which is exact.
 \item $Z^M_2 \equiv R$: while in the MSSM this is imposed to avoid the $B-L$ violating terms in the Superpotential, in the E$_6$SSM it is automatic due to the $U(1)_N$ presence. (As usual, the states which are odd under $R$-parity are called Superpartners, with the lightest Superpartner, i.e., the Lightest Supersymmetric Particle (LSP) being stable.)
 \end{itemize}

In this case, one finds that the low energy effective Superpotential is given by 
\be 
W = Y_{u} Q U^c H_u + Y_{d} Q D^c H_d + Y_{e} L E^c H_d + Y_{\nu} L \nu^c H_u  + \lambda S H_d H_u,
\label{Superpot}
\ee
where $\lambda S H_1 H_2$ stands for $\lambda_{ijk} S_i H_{d_j} H_{u_k}$. Therefore, the effective $\mu$-parameter is given by $\lambda_{333}s/\sqrt{2}$, generating the term $\mu H_{d3}H_{u3}$ in the Superpotential, thereby avoiding the so-called $\mu$-problem of the MSSM.

\begin{table}
 \begin{tabular}{|r|c|c|c|c|}
  \hline
   & $Z^H_2$ & $Z^L_2$ & $Z^B_2$ & $Z^M_2$  \\
   \hline
  $S_\alpha$ & - & + & + & +  \\
  $H_{d\alpha}, H_{u\alpha}$ & - & + & + & + \\
  $S_3$ & + & + & + & +  \\
  $H_{d3}, H_{u3}$ & + & + & + & +  \\
  $Q_{i}, u^c_{i},d^c_i$ & - & - & - & -  \\
  $L_{i}, e^c_{i}$ & - & - & - & -  \\
  $\bar{D}_i, D_{i}$ & - & + & - & +  \\
  \hline
 \end{tabular}
 \caption{Discrete symmetries in the E$_6$SSM.}\label{symmetries}
\end{table}

Before proceeding further by describing the gaugino sector of the E$_6$SSM, we shall now make a remark concerning the $Z^{\prime}$ mass bounds. In models with extra $U(1)$ gauge groups, kinetic mixing between the gauge bosons is allowed. Such a  mixing is expected to be generated through loops in the Renormalisation Group Equation (RGE) evolution \cite{Aguila1988,Aguila1988a}. The mixing opens up the decay channel $Z^{\prime}\rightarrow W^{+}W^{-}$, which easily becomes the dominant one. The increased width reduces the sensitivity of conventional resonance searches \cite{Accomando2013,Accomando2020}. Also the decays to Superpartners reduce the dilepton Branching Ratio (BR) so the usual dilepton bound of $4.5$~TeV \cite{Aad2019} becomes $3.3$~TeV. In such conditions, then, the bound from $Z^{\prime}\rightarrow W^{+}W^{-}$ is practically the same as from the dilepton searches. Our bounds are slightly lower than in \cite{Frank2020}, though, as in our case the inert Superpartners are light and take a share of the $Z'$ BR. Notice that, 
in what follows the $Z^{\prime}$ mass is relevant in the DD rates of inert neutralinos, but the corresponding results can easily be scaled to a given $Z'$ mass. In constrast, outside resonant regions,  the actual $M_{Z'}$ value does not affect the relic densities.

\subsection{Active and inert neutralino states}
Let us now consider the active sector in the E$_6$SSM. In this model, the neutralinos
$\tilde{\chi}^0$ ($i = 1,.., 6$) are the physical (mass) superpositions of three fermionic
partners of the neutral gauge bosons, called gauginos  $\tilde{B}$ (bino), $\tilde{W}^ 3$ (wino) and $\tilde{B}'$ (B'ino), plus the three fermionic partners of the neutral MSSM Higgs states, called higgsinos   
$\tilde{H}_1^0$ and $\tilde{H}_2^0$. In the basis of \( \left(\lambda_{\tilde{B}}, \tilde{W}^0, \tilde{H}_d^0, \tilde{H}_u^0, \tilde{S}, \lambda_{B'}\right)$, the active neutralino mass matrix is given by  
\begin{equation} 
m_{\tilde{\chi}^0} = \left( 
\begin{array}{cccccc}
M_1 &0 & -M_Zs_Wc_\beta & M_Zs_Ws_\beta & 0  &0\\ 
0 &M_2 & M_Zc_Wc_\beta & M_Zc_Ws_\beta &0 &0\\ 
-M_Zs_Wc_\beta & M_Zc_Wc_\beta &0 &- \frac{1}{\sqrt{2}} v_s \lambda  &- \frac{1}{\sqrt{2}}\lambda vs_\beta  &m_{\lambda_{B'} \tilde{H}^0_d} \\ 
M_Zs_Ws_\beta & M_Zc_Wc_\beta &- \frac{1}{\sqrt{2}} v_s \lambda  &0 &- \frac{1}{\sqrt{2}}\lambda vc_\beta & m_{\lambda_{B'} \tilde{H}^0_u}  \\ 
0  &0 &- \frac{1}{\sqrt{2}}\lambda vs_\beta  &- \frac{1}{\sqrt{2}}\lambda vc_\beta &0 &\frac{1}{2} \sqrt{\frac{5}{2}} g_N v_s \\ 
0 &0 & m_{\tilde{H}^0_d \lambda_{B'} }  &m_{\tilde{H}^0_u \lambda_{B'}}   &\frac{1}{2} \sqrt{\frac{5}{2}} g_N v_s &M_1' \end{array} 
\right), 
 \end{equation} 
 where $c_W\equiv \cos\theta_W$, $s_W\equiv \sin\theta_W$, $c_\beta\equiv\cos\beta$, $s_\beta\equiv\cos\beta$ while $M_1$, $M_2$ and $M'_1$ are the $U(1)_Y$, $SU(2)_L$ and $U(1)_N$ soft SUSY-breaking gaugino masses, respectively. Furthemore, one has 
 \begin{align} 
m_{\lambda_{\tilde{B}}\tilde{H}_d^0} &= -\frac{1}{20} \Big(10 g_1  + 3 \sqrt{10} g_{B Y} \Big)v_1, \\ 
m_{\lambda_{\tilde{B}}\tilde{H}_u^0} &= \Big(\frac{1}{2} g_1  - \frac{1}{\sqrt{10}} g_{B Y} \Big)v_2, \\ 
m_{\tilde{H}_d^0\lambda_{B'}} &= -\frac{1}{20} \Big(10 g_{Y B}  + 3 \sqrt{10} g_N \Big)v_1, \\ 
m_{\tilde{H}_u^0\lambda_{B'}} &= \Big(\frac{1}{2} g_{Y B}  - \frac{1}{\sqrt{10}} g_N \Big)v_2. 
\end{align} 
This matrix is diagonalised through \(N\), such that  
\begin{equation} 
N^* m_{\tilde{\chi}^0} N^{\dagger} = m^{\rm diag}_{\tilde{\chi}^0}, 
\end{equation} 
with 
\begin{align} 
\lambda_{\tilde{B}} = \sum_{j}N^*_{j 1}\lambda^0_{{j}}\,, \hspace{1cm} 
\tilde{W}^0 = \sum_{j}N^*_{j 2}\lambda^0_{{j}}\,, \hspace{1cm} 
\tilde{H}_d^0 = \sum_{j}N^*_{j 3}\lambda^0_{{j}},\\ 
\tilde{H}_u^0 = \sum_{j}N^*_{j 4}\lambda^0_{{j}}\,, \hspace{1cm} 
\tilde{S} = \sum_{j}N^*_{j 5}\lambda^0_{{j}}\,, \hspace{1cm} 
\lambda_{B'} = \sum_{j}N^*_{j 6}\lambda^0_{{j}}.
\end{align} 
In these conditions, the LSP has the following decomposition:
\be 
\tilde{\chi}^0_1 = N_{11} \lambda_{\tilde{B}}+ N_{12} \tilde{W}^0 + N_{13} \tilde{H}_d^0 + N_{14} \tilde{H}_u^0 + N_{15} \tilde{S} + N_{16}\lambda_{B'}.
\ee
In addition, the mass matrix for the inert neutralinos in the basis of \( \left(\tilde{h}^{0,I}_{d1}, \tilde{h}^{0,I}_{d2}, \tilde{h}^{0,I}_{u1}, \tilde{h}^{0,I}_{u2}\right)\) is given by 
\begin{equation} 
m_{\tilde{\chi}^{0,I}} = \left( 
\begin{array}{cccc}
0 & 0 & -\frac{1}{\sqrt{2}}v_s\lambda_{311} & -\frac{1}{\sqrt{2}}v_s\lambda_{312} \\ 
0 & 0 & -\frac{1}{\sqrt{2}}v_s\lambda_{321} & -\frac{1}{\sqrt{2}}v_s\lambda_{322} \\ 
-\frac{1}{\sqrt{2}}v_s\lambda_{311} & -\frac{1}{\sqrt{2}}v_s\lambda_{312} & 0 & 0 \\ 
-\frac{1}{\sqrt{2}}v_s\lambda_{321} & -\frac{1}{\sqrt{2}}v_s\lambda_{322} & 0 & 0 
\end{array} 
\right). 
\end{equation}  
Finally, notice that, here, we have not included the inert singlinos which in this simplified model have no Yukawa interactions and thus are completely decoupled and massless.

\subsection{Active and inert (pseudo)scalar states}
In this type of E$_6$SSM model, the active neutral Higgs states $H_u$ and $H_d$ are mixed with the singlet scalar $S$. Therefore, the CP-even and CP-odd mass matrices are extended to $3\times 3$ instead of $2 \times 2$ matrices. The lightest CP-even neutral Higgs is the SM-like Higgs with mass equal to 125 GeV. The other Higgs bosons are typically heavier. 

In addition, we have neutral inert (pseudo)scalar states with the following mass matrix in the basis of $\left( h_1^{0I}, h_2^{0I*}\right)$, $\left( h_1^{0I*}, h_2^{0I}\right)$: 
\be
m_h^{0I} = \left( 
\begin{array}{cc}
m_{11}^2 ~ & ~m_{12}^2\\ \\
m_{12}^{2 T} & m_{22}^2 
\end{array} 
\right), 
\ee
where the $m_{ij}^2$ entries are given in terms if soft SUSY-breaking terms and the corresponding $D$-terms. It is remarkable that,  due to the discrete symmetry $Z_2^H$, the mass eigenstates of the inert fields respect the CP symmetry, hence, both CP-even and CP-odd states have equal masses and the complex fields $h_i$ given by their superposition are the physical states. 

Finally, the inert singlet scalars are completely decoupled, with the following mass matrix:
\be 
m_{s^I}^2 = - \frac{1}{16} g_N^2 \left( 2 v_2^2 + 3 v_1^2 - 5 v_s^2 \right) I_{2\times 2} + m_s^2 I_{2\times 2},
\ee
where $I_{2\times 2}$ is $(2 \times 2)$ unit matrix and $m_s^2$ is the soft SUSY-breaking term of the singlet scalar. We may then notice that the mass has a contribution of the form $g_{N}^{2}v_{s}^{2}$, which is the scale of the $Z^{\prime}$ mass, so the inert singlet scalars will never be light as long as the soft masses $m_s^2$ are positive.

\section{Two-component DM scenario}
\label{sec:DM}
We shall now look at different cases with two DM candidates. One of them will be stabilised by the $R$-parity while the other by the  $Z_{2}^H$ symmetry (hereafter, $Z_2$ for ease of notation). As we want both components to produce only a fraction of the observed DM relic abundance, we are directed towards the DM candidates that usually lead to underabundance, namely higgsinos and winos \cite{Profumo2004}. Both of these can be motivated to be the LSP: we have a higgsino LSP if the effective $\mu$-parameter is smaller than the smallest gaugino mass parameter and the wino is naturally the LSP in Anomaly Mediated SUSY  Breaking (AMSB) \cite{Randall1999}. From the $Z_{2}$-odd sector we thus have inert (pseudo)scalar states and inert higgsinos as the potential DM candidates. If the inert higgsino were  the LSP, we could also have both of the DM candidates from the inert sector. 

The physics of two-component DM differs from the single component case. The DM particles freeze out at temperatures close to $m/T\sim 20$. When the two DM components have different masses, the heavier one freezes out first and after its freeze-out it has a higher number density than it would have under thermal equilibrium. If there are annihilation processes where the lighter DM particle can coannihilate with the heavier one, these processes can be largely enhanced compared to standard freeze-out and the relic abundance can be different from the corresponding single component case by several orders of magnitude \cite{Bhattacharya2013}.

Hereafter, we define two-component DM to mean a case where both DM candidates give a sizable fraction of the total relic density. In such a case, the phenomenology could differ from a single component case and both components might be detectable. Unfortunately, this rules out inert (pseudo)scalar fields as DM candidates. They are complex scalar fields and have a coupling to the $Z$-boson and hence would have been detected in DD experiments already.
We may also note that, in the case of $m_1+m_2 > m_3$, where $m_1$ and $m_2$ are the two lightest amongst the lightest inert (pseudo)scalar,  lightest inert neutralino and  lightest active neutralino while $m_3$ is the heaviest amongst them, the latter also becomes stable despite not been protected by any discrete symmetry. For instance, let the inert (pseudo)scalar be heavier than the other two. For it to decay, the only $Z_2$-odd particle available is the inert neutralino, but such a decay would violate $R$-parity unless there is also a $Z_2$-even neutralino in the final state. As the inert scalar is excluded as DM candidate by DD  constraints (unless the relic abundance is very small), we exclude the part of parameter space which leads to three stable DM candidates.

The model files for our studies were prepared with {\tt SARAH (v4.14.1)} \cite{Staub2014}, the spectrum was extracted from {\tt SPheno (v4.0.3)} \cite{Porod2012} while the DM observables were computed with {\tt micrOmegas (v5.0.8)} \cite{Belanger2010,Belanger2015}.

\subsection{Two higgsinos as DM candidates}
A MSSM higgsino with a mass below $1$~TeV leads to underabundance in the case of a single DM component \cite{Profumo2004} and the relic density increases almost linearly with the higgsino mass. As the main annihilation channel for higgsinos is neutralino-chargino coannihilations via SM gauge bosons, the inert neutralino should behave similarly. Hence sub-TeV higgsinos would be potential DM candidates in a two-component scenario. This is what we also find when we scan the parameter space. 

\begin{figure}
\begin{center}
\includegraphics[width=13cm, height=7.5cm]{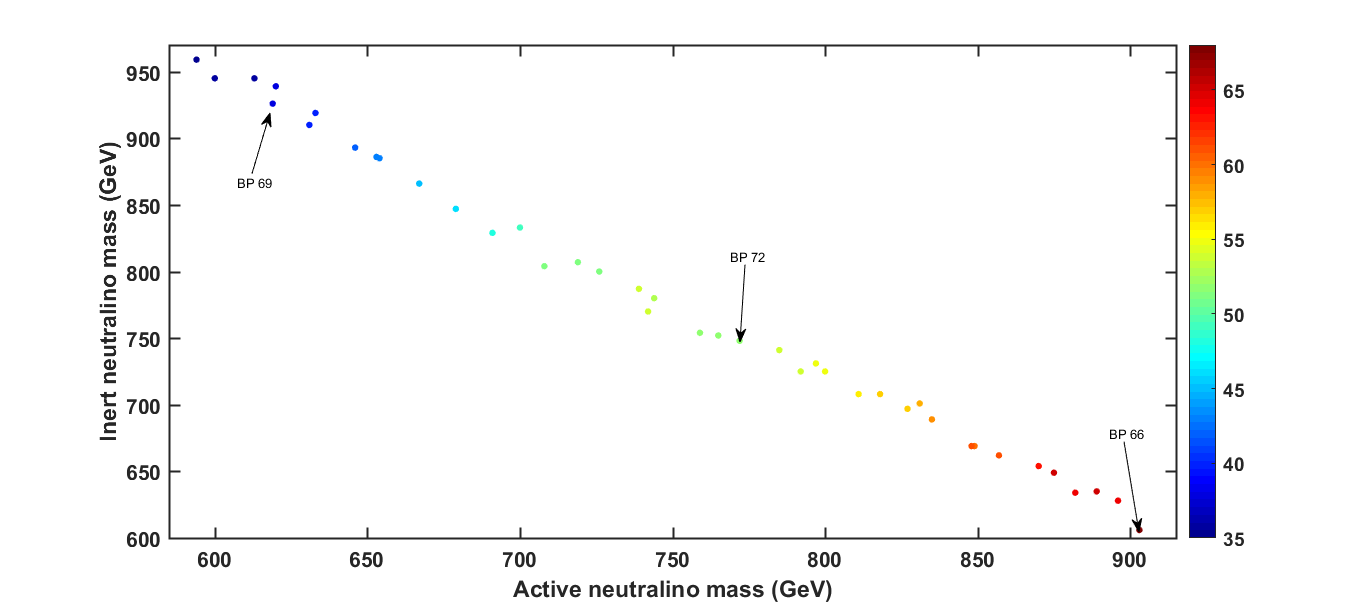}
\end{center}
\caption{The data points that give a relic density of $\Omega_{\mathrm{CDM}}h^{2}=0.120\pm 0.002$. The color indicates the percentage of the active higgsino component of the total relic density. We also pick three of the data points, indicated by the arrows, for studying DD and ID rates in more detail.\label{fig:2higgsinorelic}}
\end{figure}

\begin{figure}
\begin{center}
\includegraphics[width=13cm, height=7.5cm]{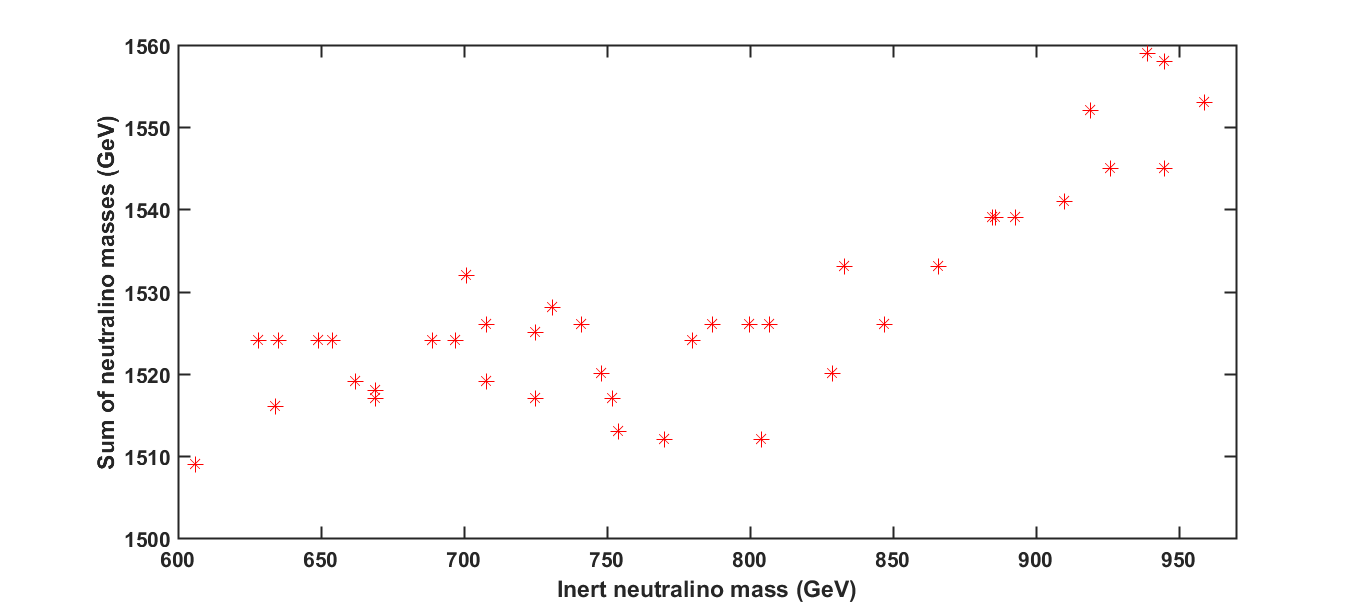}
\end{center}
\caption{The sum of the two neutralino masses as a function of the inert higgsino mass. We see that the sum of the masses is larger  when the inert neutralino is the heavier of the two.\label{fig:masssum}}
\end{figure}

The two higgsinos annihilate nearly independently, so the sum of their masses is nearly constant when we require $\Omega_{\mathrm{CDM}}h^{2}=0.120\pm 0.002$. We plot the viable data points in Fig.~\ref{fig:2higgsinorelic} and indicate by colour the fraction of the active higgsino component to the total relic density. For these points
\begin{equation}\label{eq:masssum}
m_{\tilde{H}^{0}}+m_{\tilde{H}^{0,I}}=1.53\pm0.03\;\mathrm{TeV}.
\end{equation}
The sum of the masses is at the upper half of this interval if the inert neutralino is heavier and at the lower half of this interval if the active neutralino is heavier,  as show in Fig. \ref{fig:masssum}. We shall discuss the reasons for this below.

Outside the region shown in Fig. \ref{fig:masssum} the sum of the two neutralino masses varies relatively smoothly and stays within the interval given in equation (\ref{eq:masssum}) if the active neutralino mass is 400~GeV to 1000~GeV and the inert neutralino mass is 500~GeV to 1100~GeV. Outside this interval the sum is below $1.5$~TeV. The highest possible masses are $1150$~GeV for the active neutralino and $1200$~GeV for the inert neutralino in which cases the single component saturates the relic density bound by itself.


In our scan we have kept the other particles so heavy that resonant annihilation does not occur. If, however, it happened to be that one of the DM candidates had a mass close to, say, $M_{Z^{\prime}}/2$, the resonant annihilation could give the correct relic density in a configuration that would otherwise lead to overabundance.
For further examination, we pick Benchmark Points (BPs) with neutralino masses given in Tab. \ref{tb:benchmarks}. One of the BPs has a heavy active and a light inert higgsino, one has a heavy inert and a light active higgsino and one has roughly degenerate DM candidates. For all of the BPs we have $M_{Z^{\prime}}\simeq 3.3$~TeV, which is the experimental lower bound in our case.

\begin{table}
\begin{tabular}{l c c c c}
\hline
\hline
Benchmark & Active mass & Inert mass & $\Omega^{\mathrm{A}}h^{2}$ & $\Omega^{\mathrm{I}}h^{2}$\\
\hline
BP66 & $903$ & $606$ & $0.0804$ & $0.0382$\\
BP69 & $619$ & $926$ & $0.0461$ & $0.0731$\\
BP72 & $766$ & $753$ & $0.0637$ & $0.0575$\\
\hline
\hline
\end{tabular}
\caption{The BPs chosen for further investigation. The masses are given in GeV's.\label{tb:benchmarks}}
\end{table}

\begin{figure}
\begin{center}
\includegraphics[width=13cm, height=7.5cm]{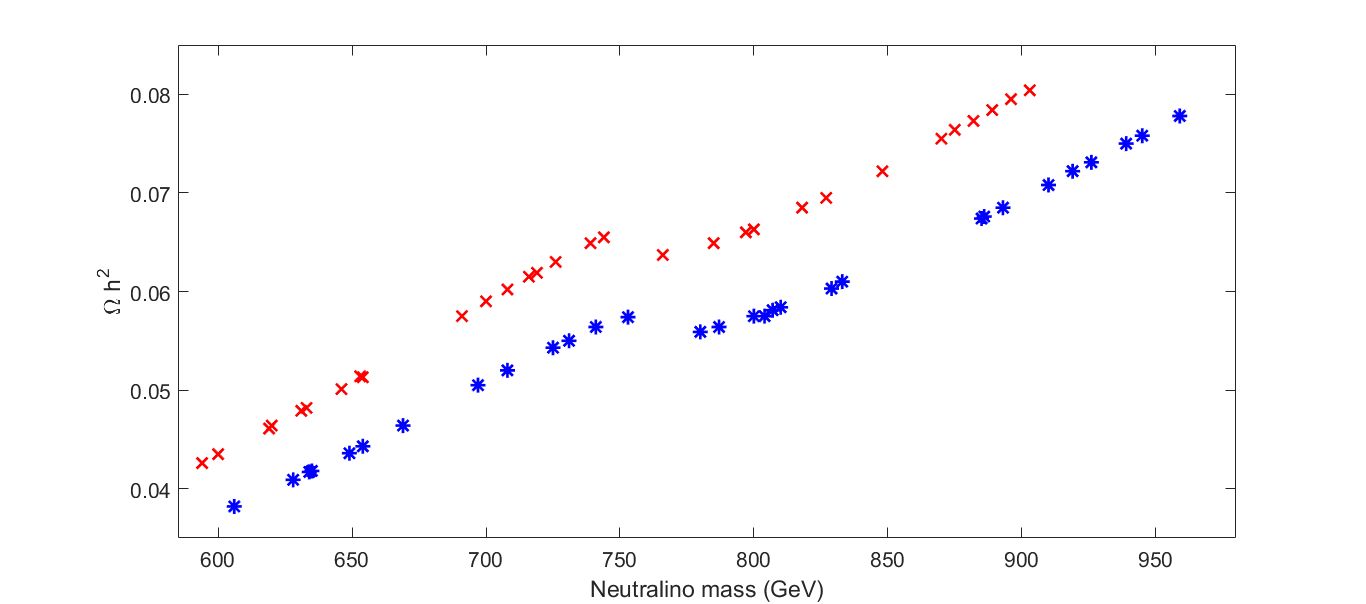}
\end{center}
\caption{Relic density contributions of the individual higgsino components (red for active higgsino, blue for inert higgsino) for the data points that satisfy the relic density constraint. The difference between the inert and active components is due to the larger mass splitting of the active neutralino and chargino.\label{fig:higgsinorelics}}
\end{figure}

We show the relic densities of the individual higgsino components for the points satisfying the relic density constraint as a function of their masses in Fig.~\ref{fig:higgsinorelics}. There is a small difference in the relic density, which is due to the larger Boltzmann suppression (the mass splittings are larger) in the case of the active higgsino, which reduces the neutralino-chargino coannihilation rate. We can also see that the heavier component has a slightly smaller relic density than it would have had in a single DM component scenario. This is due to charged current interactions, where the lighter chargino scatters from the heavier neutralino and produces a heavy chargino and a light neutralino and the heavy chargino then annihilates a heavy neutralino. This charged current process is more efficient when the colliding particles have a similar mass so that the lab frame is close to the center-of-mass frame. For SM particles the situation would be close to a fixed target scattering, where the threshold energy for chargino production is larger.

This process also explains why the sum of masses is slightly different for different mass orderings. The mass splitting between the active chargino and active neutralino is larger so that $\tilde{\chi}^{\pm}\tilde{\chi}^{0I}\rightarrow \tilde{\chi}^{0}\tilde{\chi}^{\pm I}$ is always kinematically allowed whereas there is a threshold for the process $\tilde{\chi}^{\pm I}\tilde{\chi}^{0}\rightarrow \tilde{\chi}^{0 I}\tilde{\chi}^{\pm}$. Hence the effect of the lighter DM component is larger in the case when the active one is lighter and hence the sum of masses also is.

The annihilation proceeds mostly through the SM gauge bosons either as neutralino-neutralino annihilation through the $Z$ boson or as neutralino-chargino coannihilation through the $W^{\pm}$ so the annihilation cross section is almost completely insensitive to scanning parameters besides the higgsino masses. The only other annihilation channel that could contribute is via a  Higgs boson through the Superpotential coupling $\lambda SH_{u}H_{d}$, but that only contributes through the singlino component of the higgsino, which is always tiny. Furthermore, the coupling $\lambda$ is $\mathcal{O}(0.1)$, {i.e.}, clearly smaller than the gauge couplings. Hence, a Higgs mediated contribution is always below the percent level.

The scenario with two higgsinos is also viable from the viewpoint of DD bounds. As we discuss in the next section, the Spin-Independent (SI) cross section for the active higgsino is about an order of magnitude below the limits from Xenon1T \cite{Aprile2018} while for the inert higgsino the scattering cross section is a couple of orders smaller than for the active one. Also the Spin-Dependent (SD) cross section is larger for the active higgsino.

\subsection{Wino and inert higgsino as DM candidates}
The wino as a DM candidate leads to underabundance of relic DM, if it is lighter than $2$~TeV \cite{Profumo2004}. However, together with the inert higgsino, one may achieve the correct relic density. Also in this case the annihilations are basically independent. Due to the more efficient annihilation process of the wino, the masses of the two neutralinos need to be larger than in the case of two higgsinos. Typically, the wino needs to be twice as heavy as the higgsino to get the same relic density. The wino component gives roughly $30\%$ of the relic density when the two DM candidates are degenerate, which happens around $950$~GeV.

As the abundance of the wino component is smaller and it couples to the $Z$ boson only through its mixing with the higgsinos, this scenario will be harder to discover through DD experiments. The SI DD cross section $\sigma^{\rm SI}\Omega_{\tilde{W}}/\Omega_{tot}\simeq 10^{-10}$~pb, which might eventually be detectable (as we shall se below). As the case with two higgsinos has the higher chance for detection, we shall concentrate on it in the following.

\subsection{Other combinations}
None of the other scenarios produces a viable pair of DM candidates. In the inert sector the inert doublet scalars have too large a DD cross section, which rules these out. The inert singlets have masses of the order of the $Z^{\prime}$ mass and hence they are never the lightest particles in the inert sector.
Finally, also the non-MSSM gaugino $\tilde{B}^{\prime}$ has a mass of the order of the $Z^{\prime}$ mass and hence it will not be the lightest Superpartner while the usual bino leads to overabundance without resonant annihilation.

%
\section{Two Higgsinos DM and Direct Detection Experiments}\label{sec:DD}

\begin{figure}
\begin{center}
\includegraphics[width=0.85\textwidth]{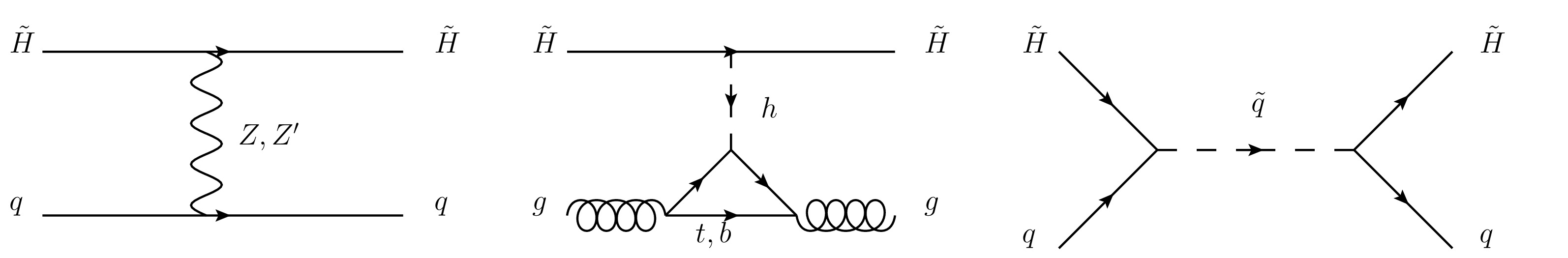}
\end{center}
\caption{The Feynman diagrams contributing to the DD of higgsino DM in the E$_6$SSM. For the inert higgsino only the $Z^{\prime}$ channel comes with unsuppressed couplings.\label{fig:higgsinodd}}
\end{figure}

We now discuss the SI and SD DM scattering cross section of the active and inert higgsino DM that we discussed in the previous section. The relevant Feynman diagrams for DD are given in Fig. \ref{fig:higgsinodd}.

The effective scalar interactions of these two DM candidates with up and down quarks are mainly given by $Z'$ exchange and the interaction with gluons is induced by Higgs exchange through heavy-quark loops, {i.e.},  
\be%
{\cal L_{\text{eff}}} = f_q \overline{\tilde{\chi}} \tilde{\chi} \, \bar{q} q+b\alpha_{s}\overline{\tilde{\chi}} \tilde{\chi}G^{\mu\nu a}G^{a}_{\mu\nu}, %
\label{scalar}
\ee%
where $f_q \propto g^2_{N}/M_{Z'}^2$ comes from the $Z^{\prime}$ mediated process and $b$ is Higgs-gluon coupling induced by the heavy quark loops. 
The effective coupling of $\tilde{\chi}$  with protons and neutrons $f_p$, $f_n$ can be computed in terms of $f_u$, $f_d$ and $b$  \cite{Jungman1996} with 
the zero momentum transfer scalar cross section of the higgsino scattering with the nucleus given by \cite{Belanger2009}:
\be%
\sigma^{\rm SI}_0 = \frac{4 m_r^2}{\pi} \left(Z f_p + (A-Z) f_n \right)^2,
\ee%
where $Z$ and $A-Z$ are the number of protons and neutrons, respectively, $m_r=m_N m_{\tilde{\chi}_1}/(m_N+m_{\tilde{\chi}_1})$, where $m_N$ is the nucleus mass. Thus, the differential scalar cross section for non-zero momentum transfer $q$ can be written as
\be%
\frac{d \sigma_{\rm SI}}{dq^2} = \frac{\sigma^{\rm SI}_0}{4 m_r^2 v^2}F^2(q^2),\; 0< q^2 < 4 m^2_r v^2,
\ee%
where $v$ is the velocity of the lightest neutralino and $F(q^2)$ is the relevant Form Factor (FF) \cite{Jungman1996}. Therefore, the SI (scattering) cross section of the LSP with a proton is given by
\be
\sigma_{\rm SI}^p=\int_0^{4 m^2_r v^2}\frac{d \sigma_{\rm SI}}{dq^2}\big{|}_{f_n=f_p} dq^2.
\ee

\begin{figure}[ht]
\begin{center}
\includegraphics[width=0.475\textwidth]{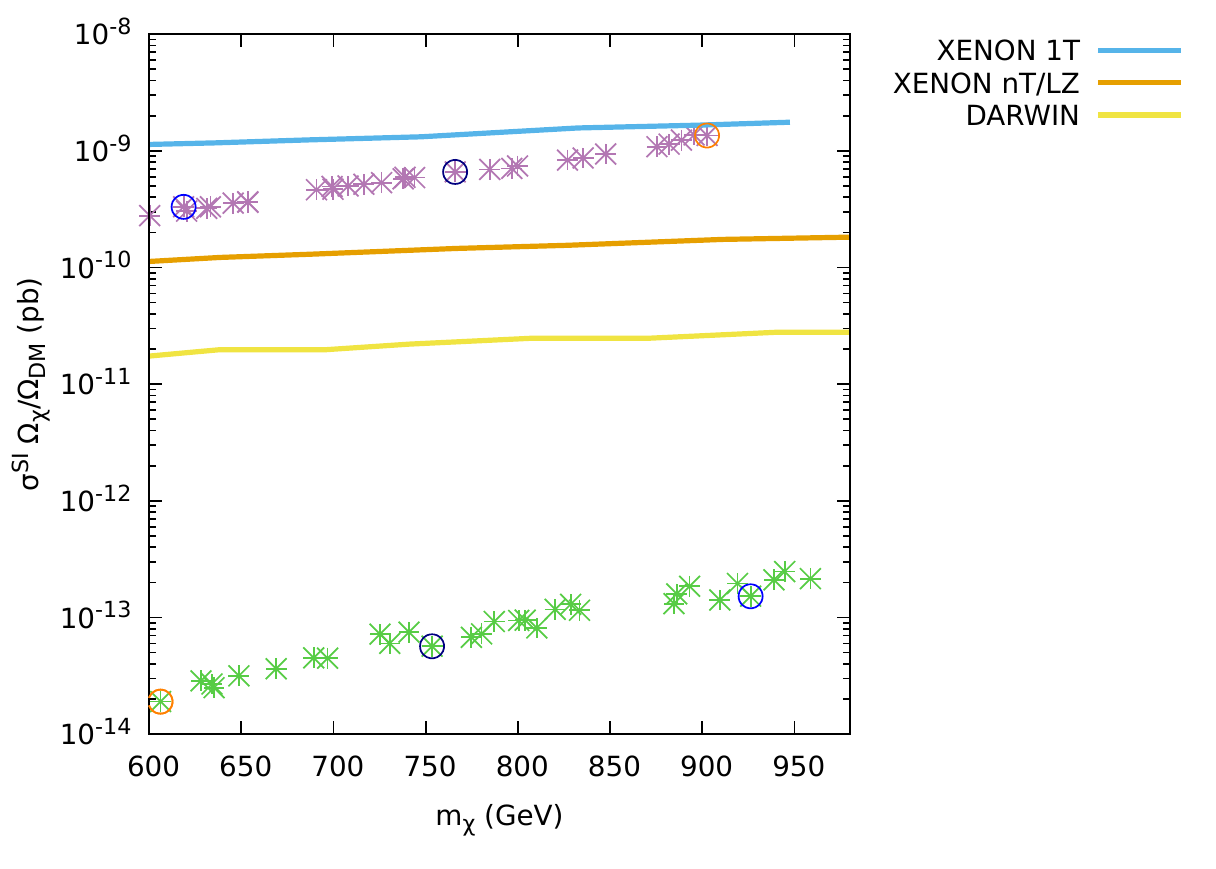}
\includegraphics[width=0.475\textwidth]{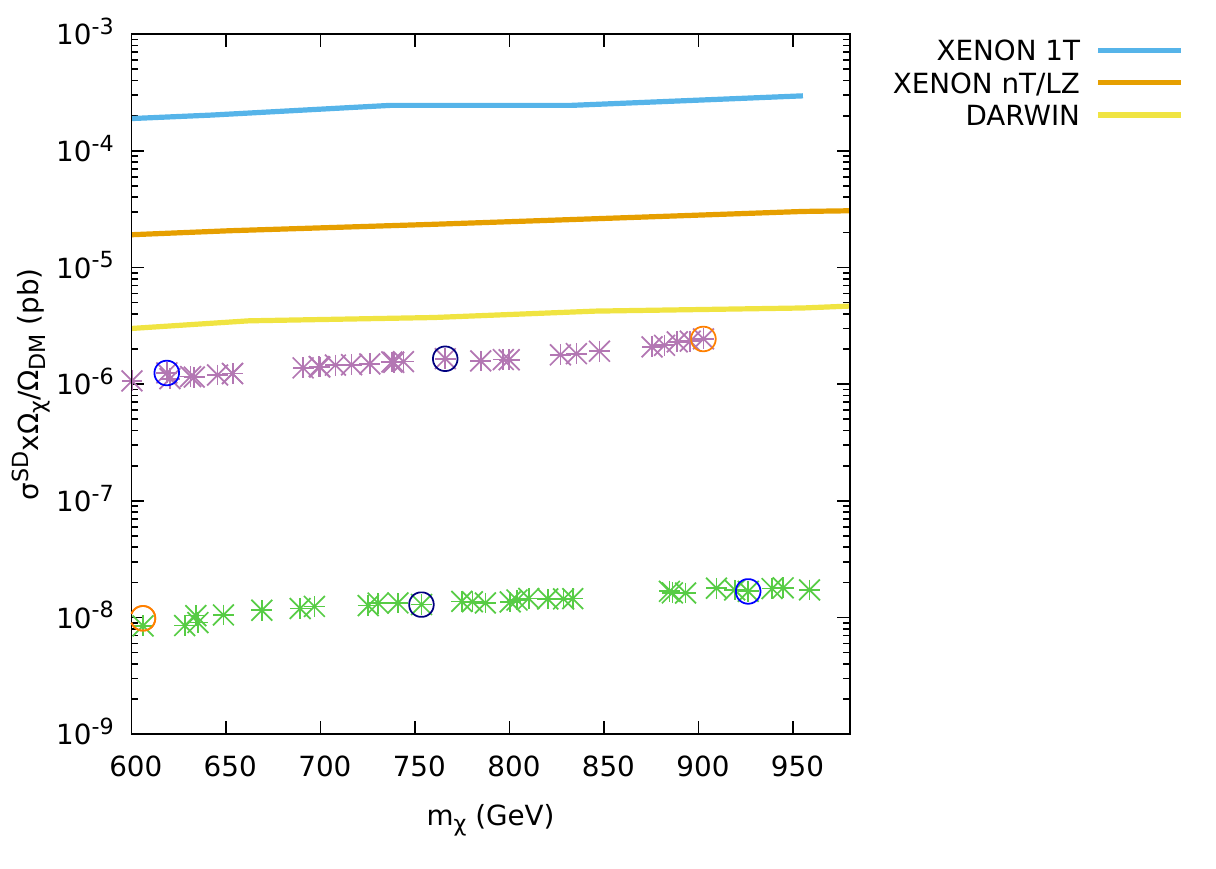}
\end{center}
\caption{SI (left) and SD (right) cross sections of active neutralino (purple) and inert neutralino (green) scattering with protons, as a function of their masses. All these points satisfy the XENON-1T exclusion region, and the active DM candidate is within the region for future DD experiments like XENON-nT and DARWIN (in here, re-scaled by the factor $\Omega_i/\Omega$) \cite{Aalbers2016}. We have highlighted the exemplary BPs 66 (orange), 69 (blue) and 72 (navy).}
\label{SI-SD}
\end{figure}


The SD interaction of a DM candidate stems solely from the quark axial current: 
$$ a_N  \bar{\chi} \gamma^\mu \gamma_5 \chi ~ \bar{N} \gamma_\mu \gamma_5 N,$$ 
where $a_N = \sum_{q=u,d,s} d_q \Delta_q^{(N)}$, with $d_q$  the effective quark level axial-vector
and pseudoscalar couplings and $\Delta_q^{(N)}$ is given via 
$\Delta_u^{(p)}=\Delta_d^{(n)}= 0.77 $, $\Delta_d^{(p)}=\Delta_u^{(n)}= -0.40 $, and $\Delta_s^{(p)}=\Delta_s^{(n)}= -0.12 $.   
In this case, the SD (scattering) cross section of DM-nucleus is given by 
\be 
\sigma_{\rm SD} = \frac{16}{\pi} m_r^2 a_N^2 J_N (J_N +1), 
\ee
where $J_N$ is the angular momentum of the target nucleus. In case of the proton target, $J_N =1/2$.

In Fig. \ref{SI-SD}, we display the SI and SD  cross sections of the active (left panel) and inert (right panel) higgsino LSP with a proton after imposing the relic abundance constraints. As the DM-nucleon recoil rates are dependent on the local density of the DM candidate, in the case of multicomponent DM the density of each species is smaller and then it is necessary to re-scale the $\Omega_i/\Omega$ factor, where $\Omega_ih^2$ is the relic abundance for the active (${\tilde H}^0$) or the inert (${\tilde H}^{0,I}$) neutralino. All our BPs satisfy the XENON-1T exclusion region and, in the case of SI interactions, the active DM candidate is clearly within the region of visibility for future DD experiments like XENON-nT and DARWIN \cite{Aalbers2016}. However, the cross section of the inert higgsino is too low and therefore falls into the neutrino floor or neutrino coherent scattering \cite{Baudis2014}, where it will be challenging to probe in the future. We have highlighted the exemplary BPs 66 (orange), 69 (blue) and 72 (navy).

The differences between the active and inert neutralinos can be understood rather easily. In the SD case the higgsino coupling to the $Z$ boson comes from
\begin{equation}
\mathcal{L}=\frac{g}{4\cos\theta_{W}}\overline{\tilde{\chi}}\gamma^{\mu}\gamma^{5}(|N_{13}|^{2}-|N_{14}|^{2})\tilde{\chi}Z_{\mu},
\end{equation}
where $N_{13}$ and $N_{14}$ give the $\tilde{H}_{u}$ and $\tilde{H}_{d}$ components of the lightest neutralino. If the mass matrix  had  only the $\mu$-term, then $|N_{13}|=|N_{14}|=\frac{1}{\sqrt{2}}$ and the coupling would vanish. Since the active higgsinos mix with the gauginos, $|N_{13}|\neq|N_{14}|$ and hence the coupling is non-zero, while for inert higgsinos the coupling vanishes as they do not mix with other states. When gauge kinetic mixing is introduced, also the inert higgsino gets a coupling to the $Z$ boson but this is much smaller than that of the active higgsino.

The difference in the SI cross section arises from the SM-like Higgs mediated scattering, which again is only relevant for the active higgsino as it mixes with the singlino. The SI cross section of the inert higgsino arises through the $Z^{\prime}$, the active singlet (which are heavy) or the tiny singlet-doublet mixing of the SM-like Higgs. Also, the squarks can act as mediators in the case of an active higgsino, but their contribution is so small that it can be neglected. 
For the inert higgsino the $Z^{\prime}$ is the only mediator with unsuppressed couplings. For the $Z$ channel the coupling is suppressed by the small kinetic mixing and for scalar mediators by the even smaller singlet-doublet mixing. As mentioned before, for these BPs, we have taken $M_{Z^{\prime}}\simeq 3.3$~TeV and  used $g_{N}=0.41$. The DD cross sections arising from $Z^{\prime}$ for other masses and couplings will scale as $g_{N}^{4}/M_{Z^{\prime}}^{4}$.



\begin{figure}
 \includegraphics[scale=0.4]{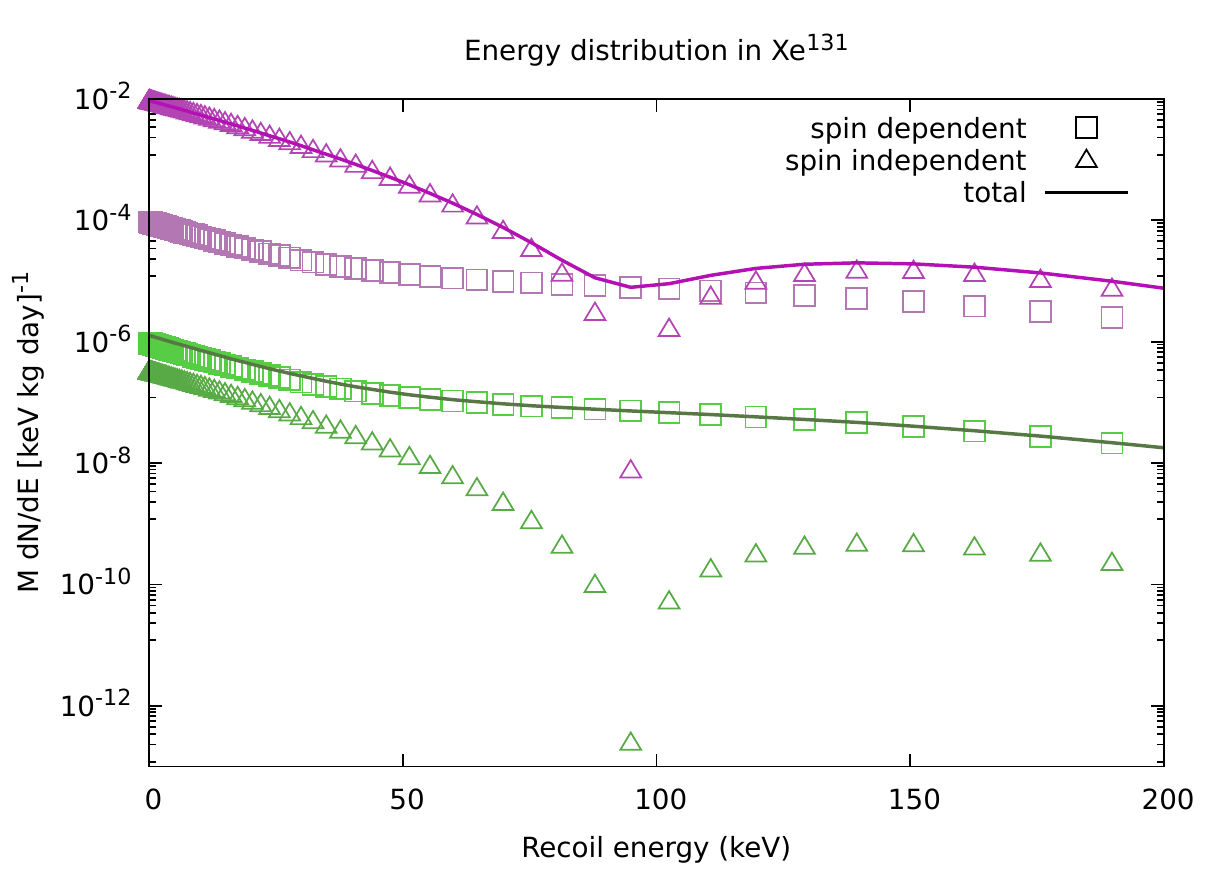}
 \includegraphics[scale=0.4]{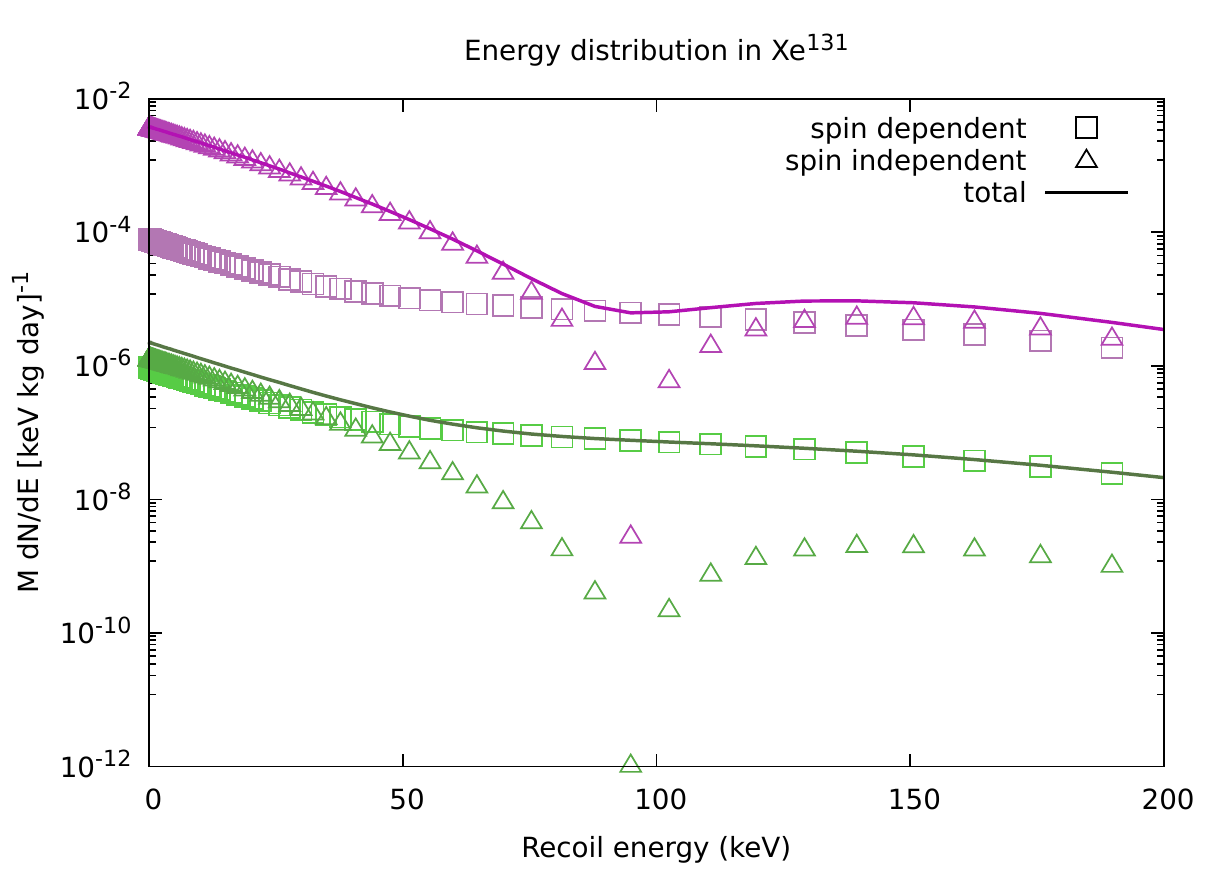}
 \includegraphics[scale=0.4]{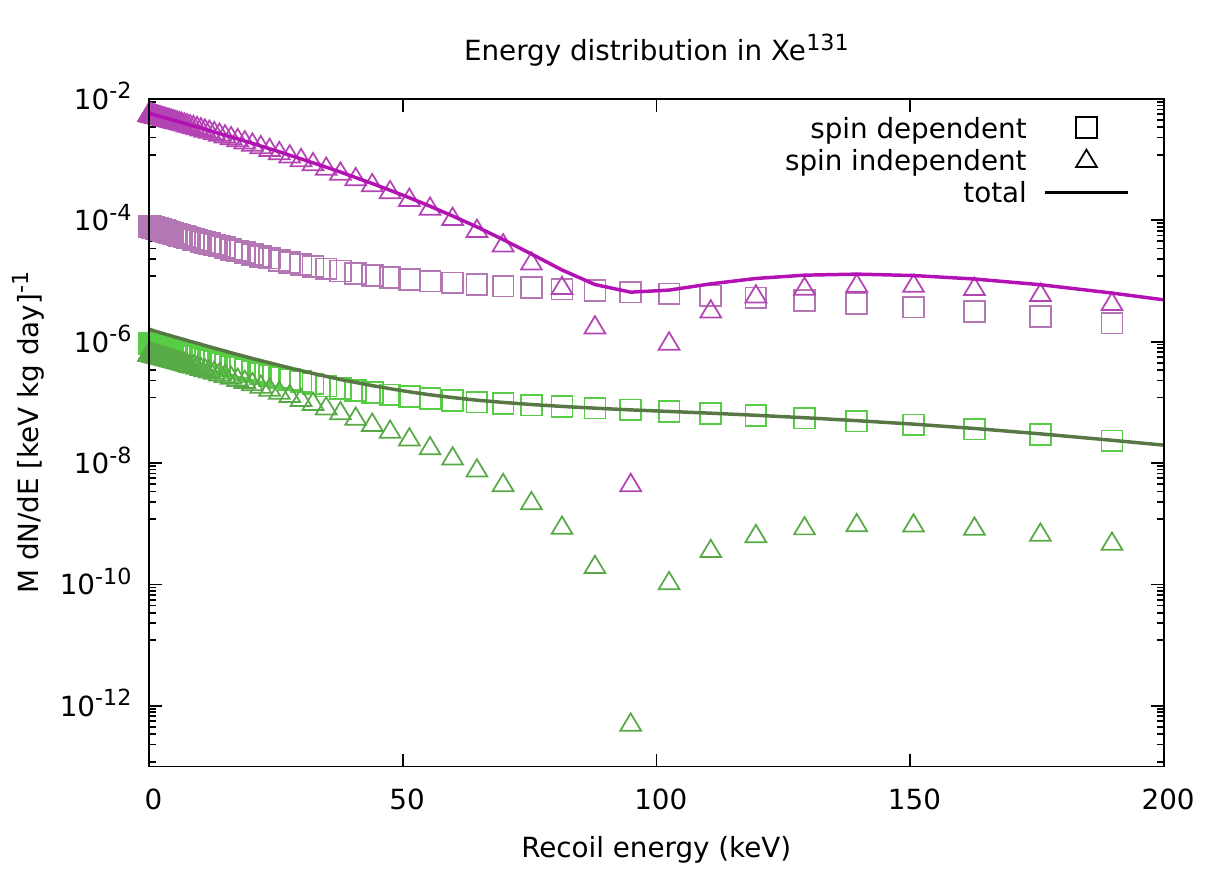}
 \includegraphics[scale=0.4]{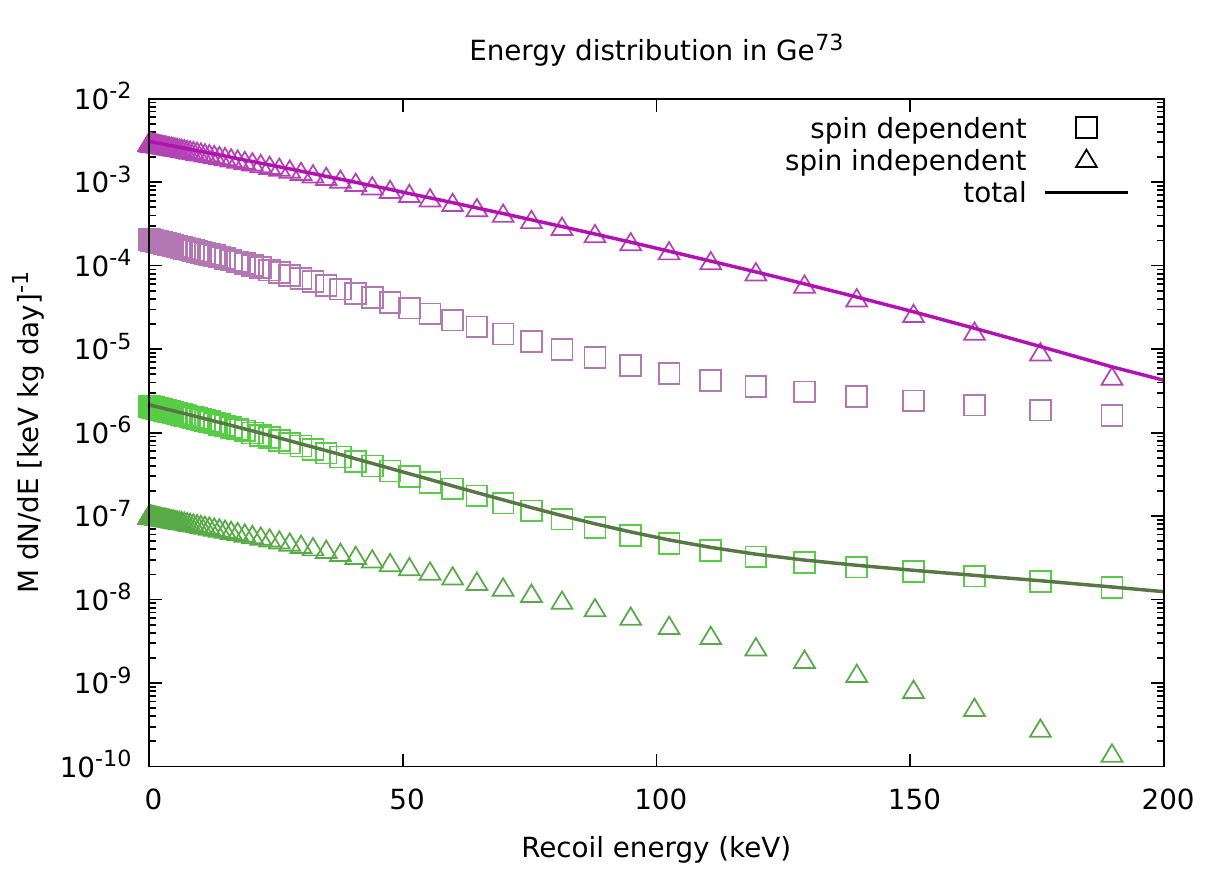}
 \includegraphics[scale=0.4]{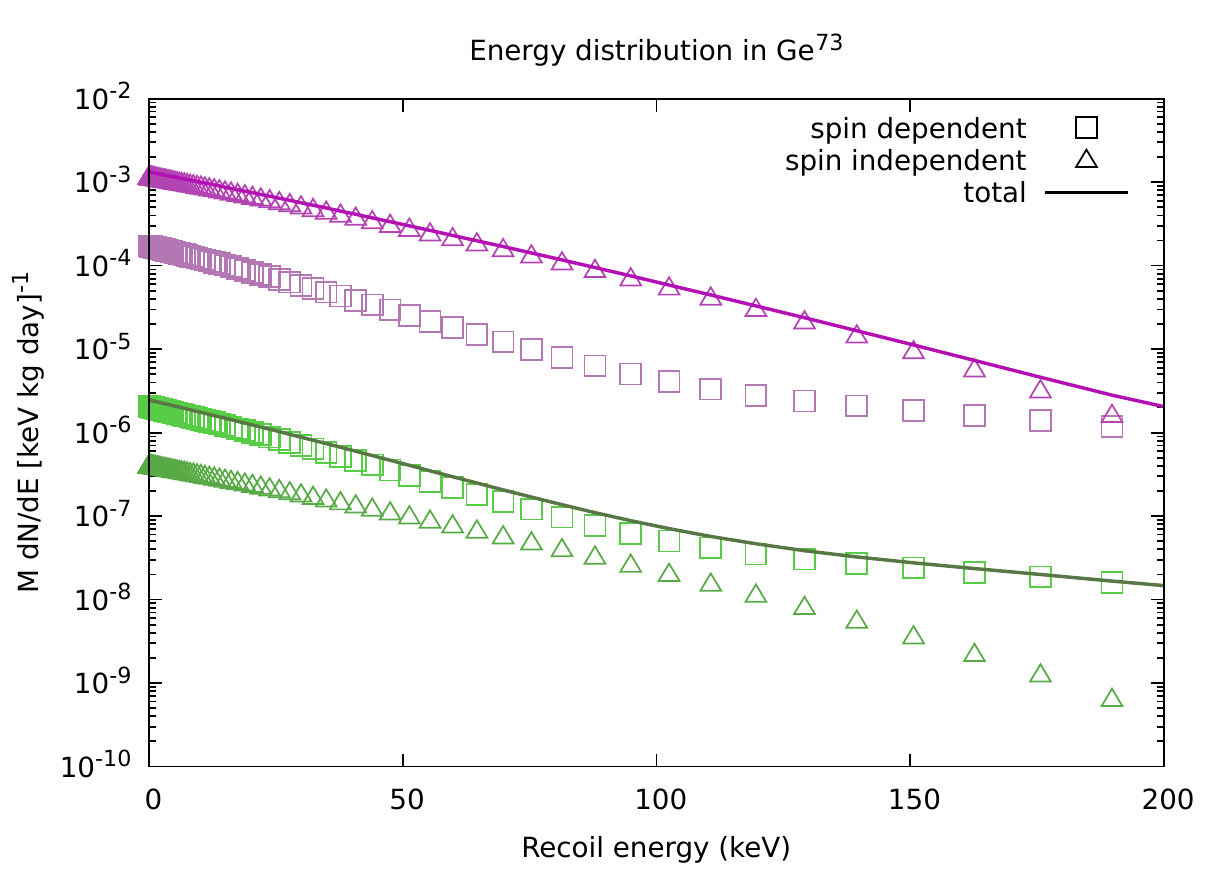}
 \includegraphics[scale=0.4]{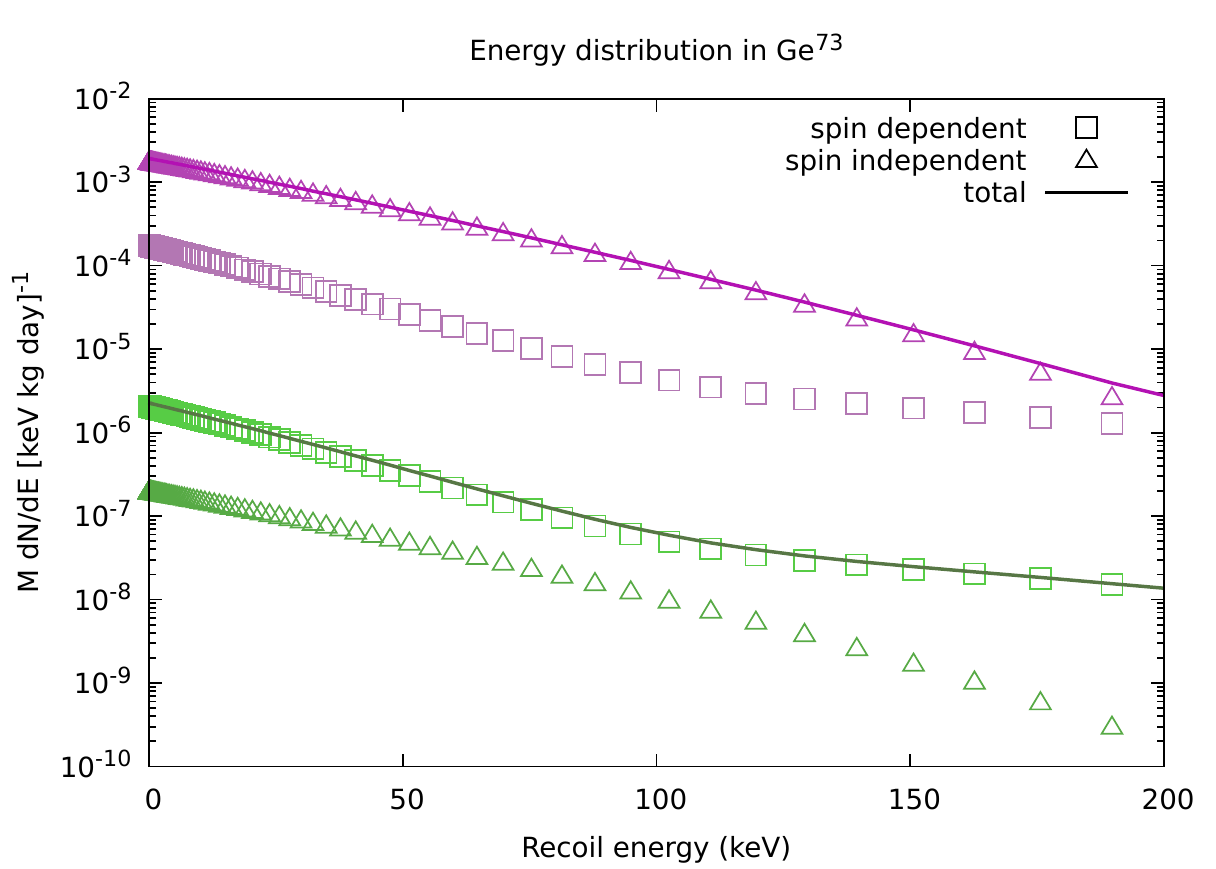}
 \includegraphics[scale=0.4]{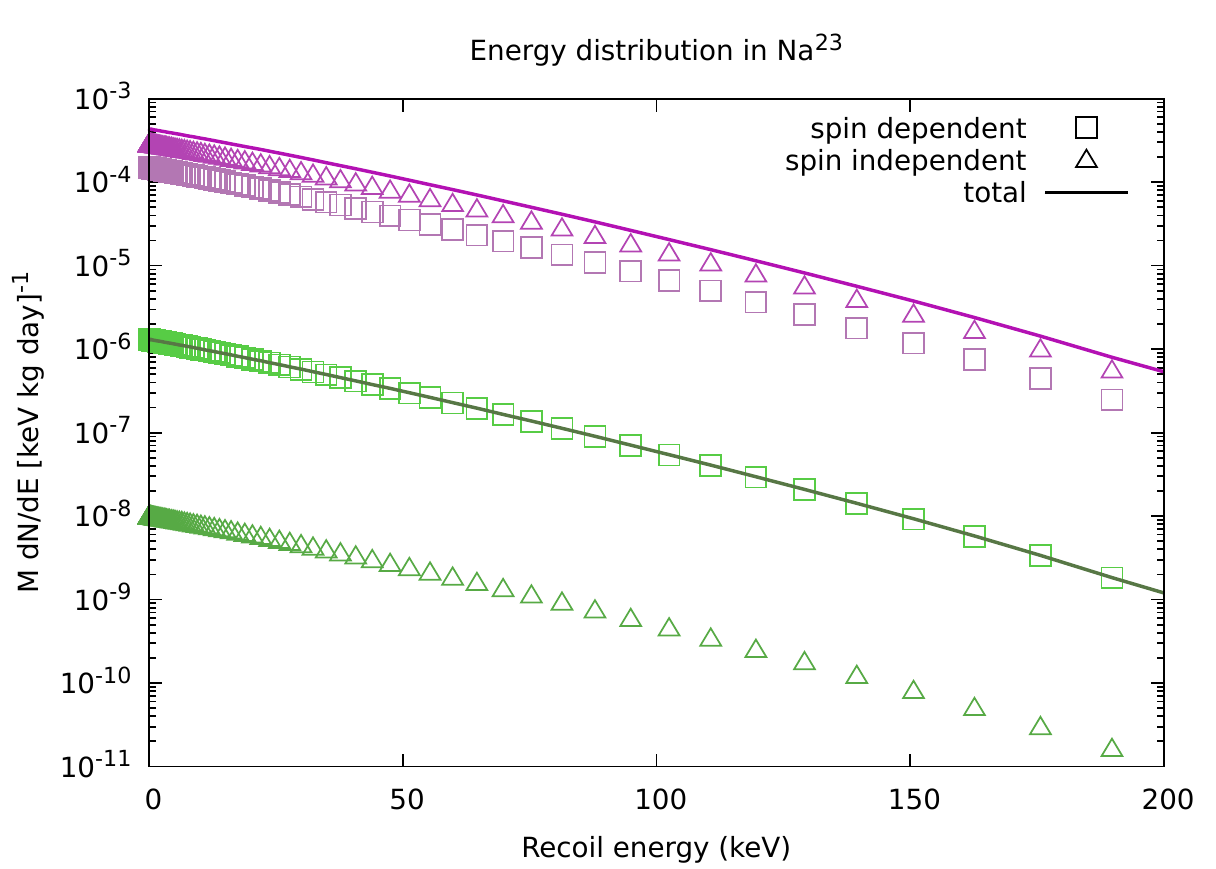}
 \includegraphics[scale=0.4]{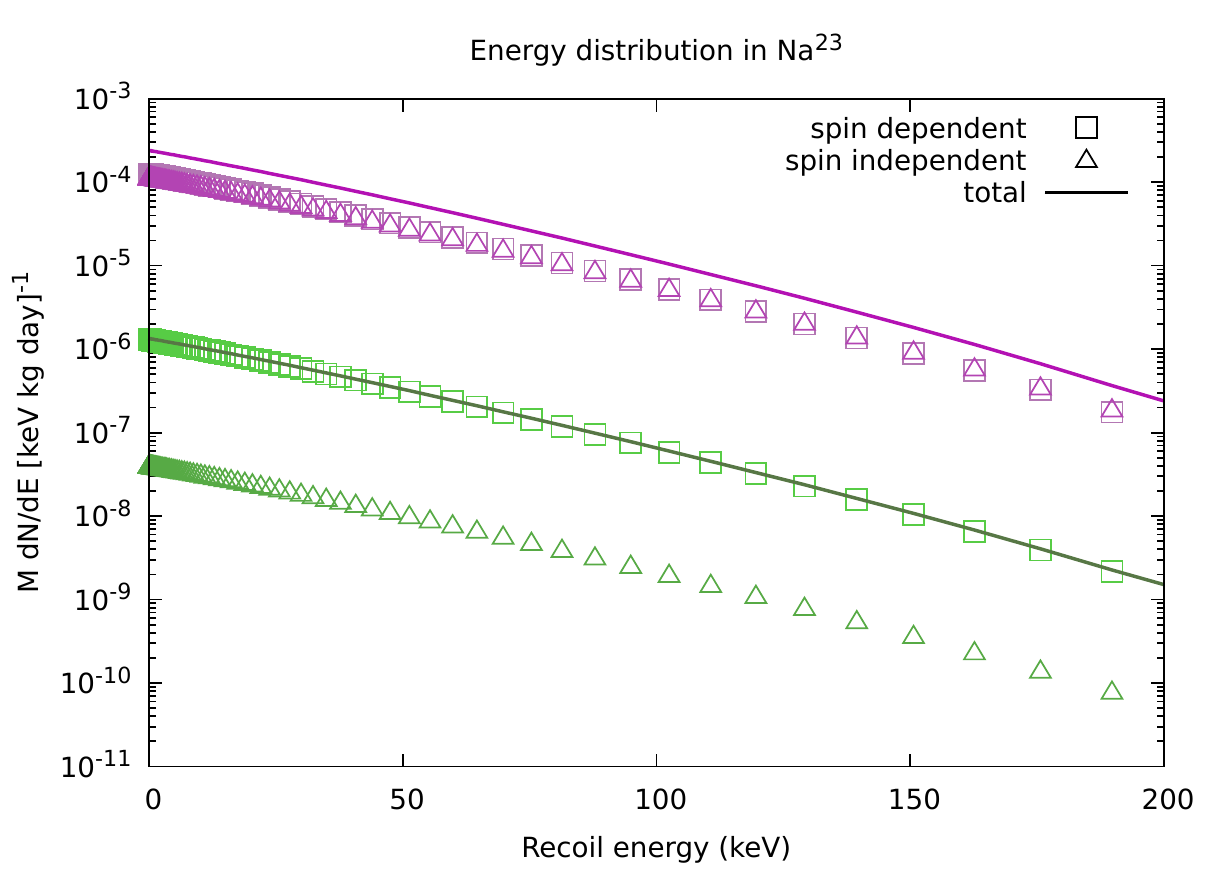} 
 \includegraphics[scale=0.4]{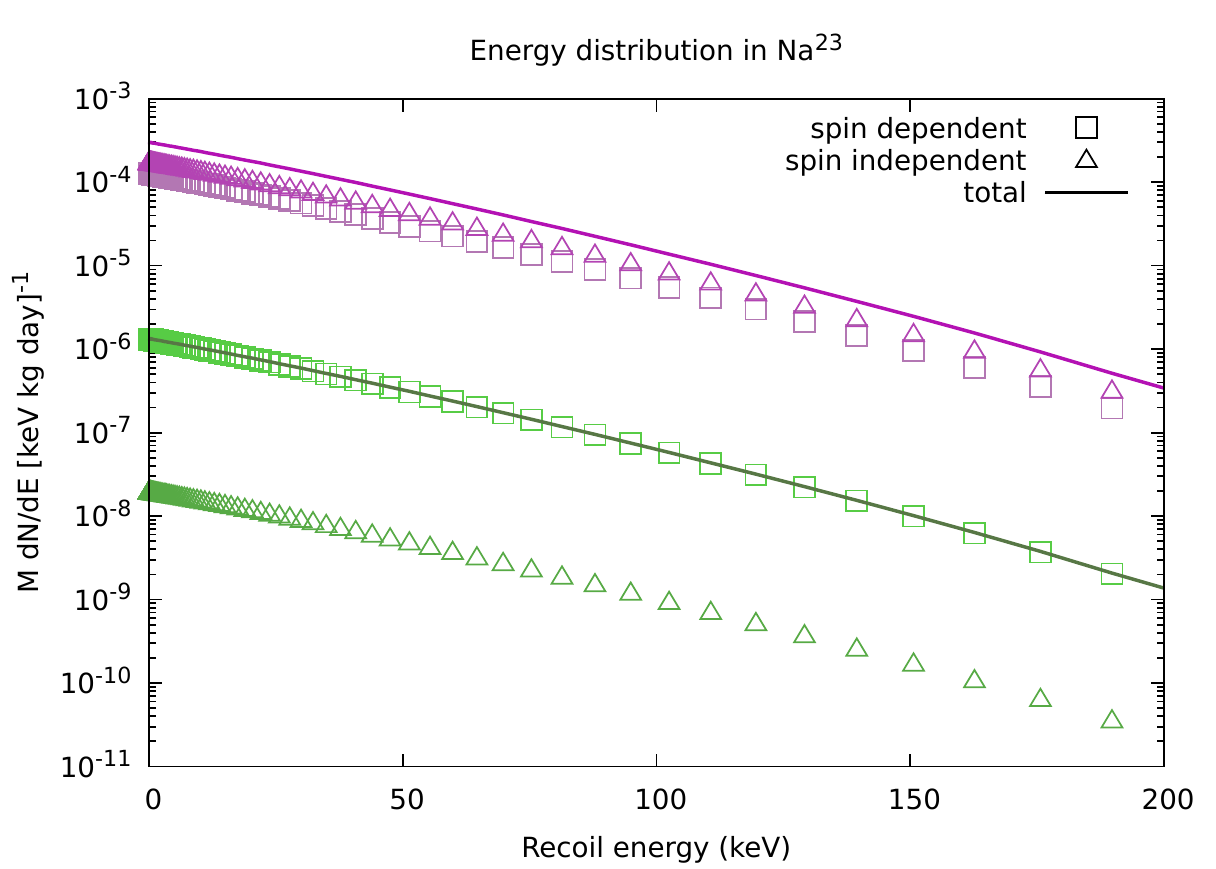}
 \includegraphics[scale=0.4]{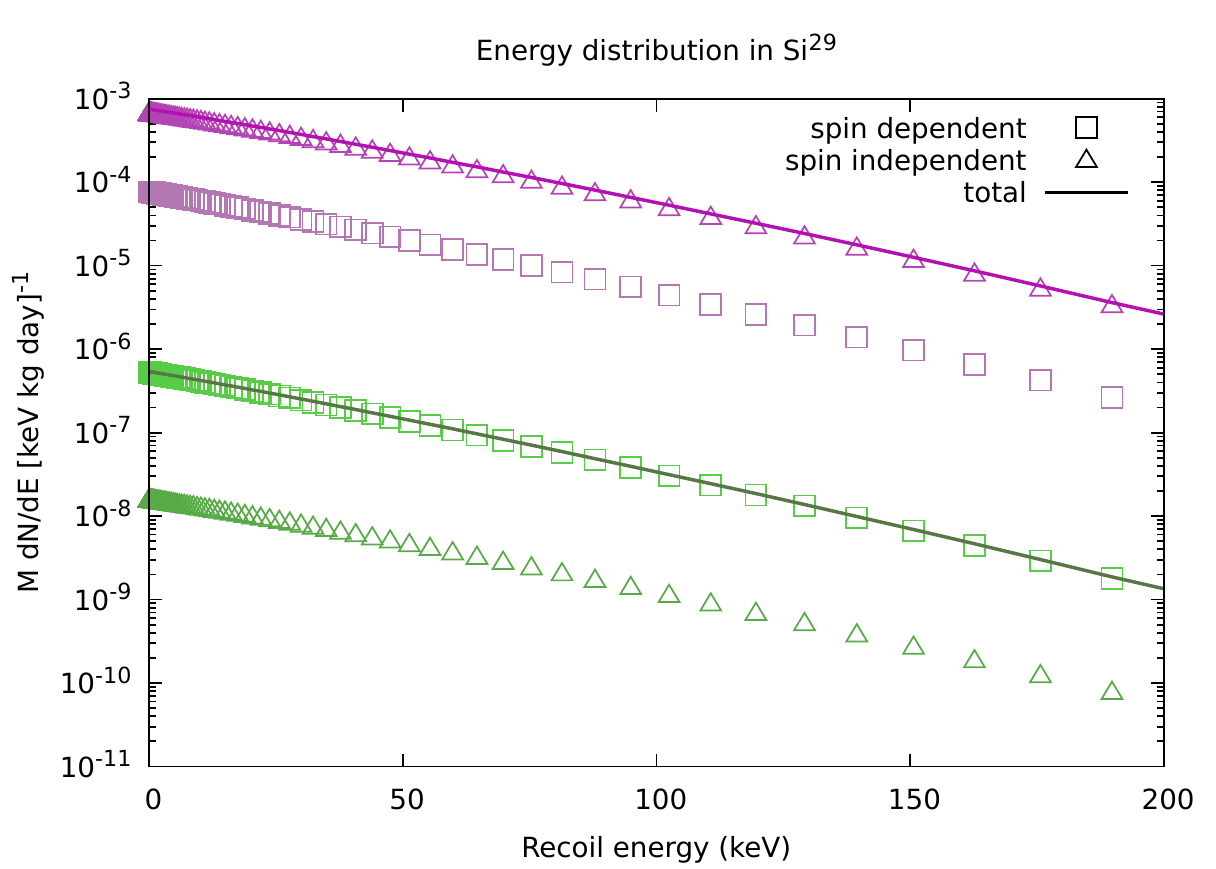}
 \includegraphics[scale=0.4]{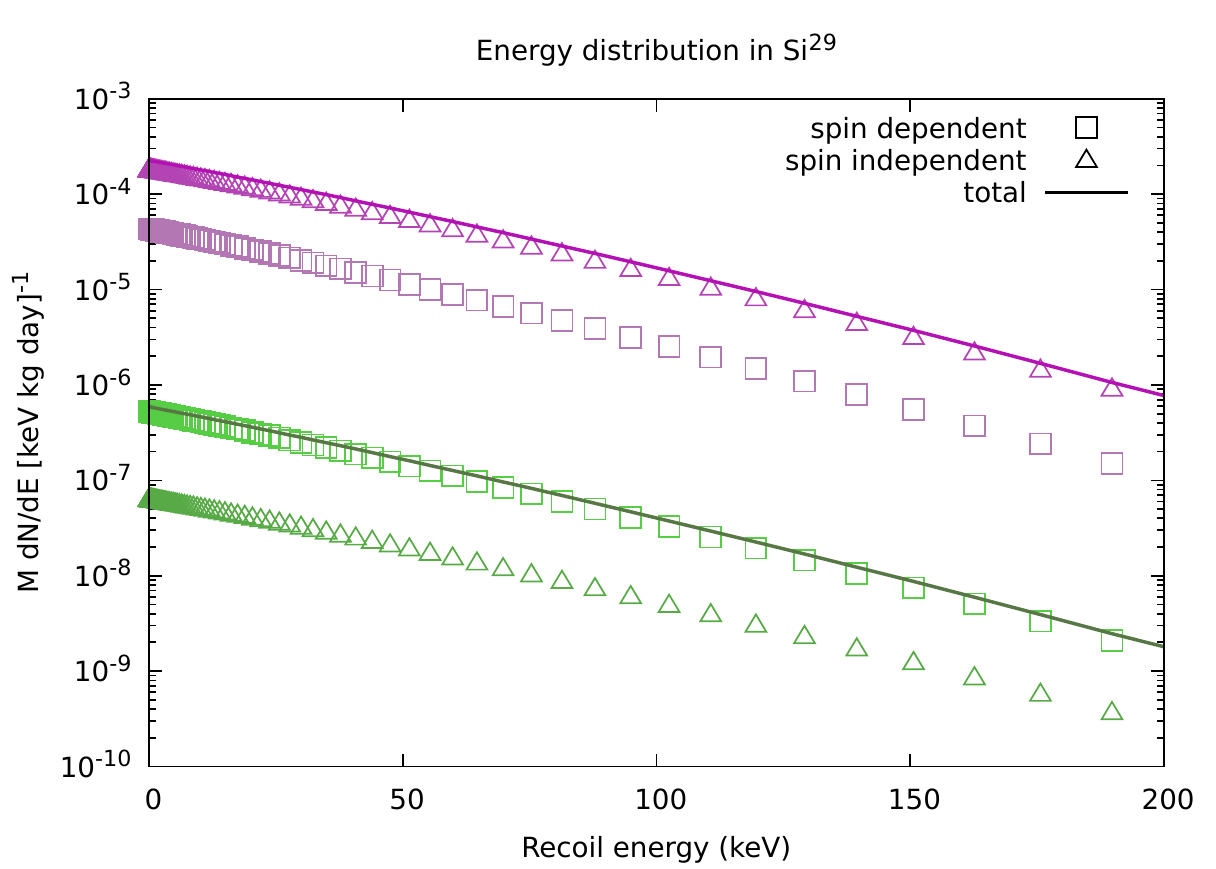} 
 \includegraphics[scale=0.4]{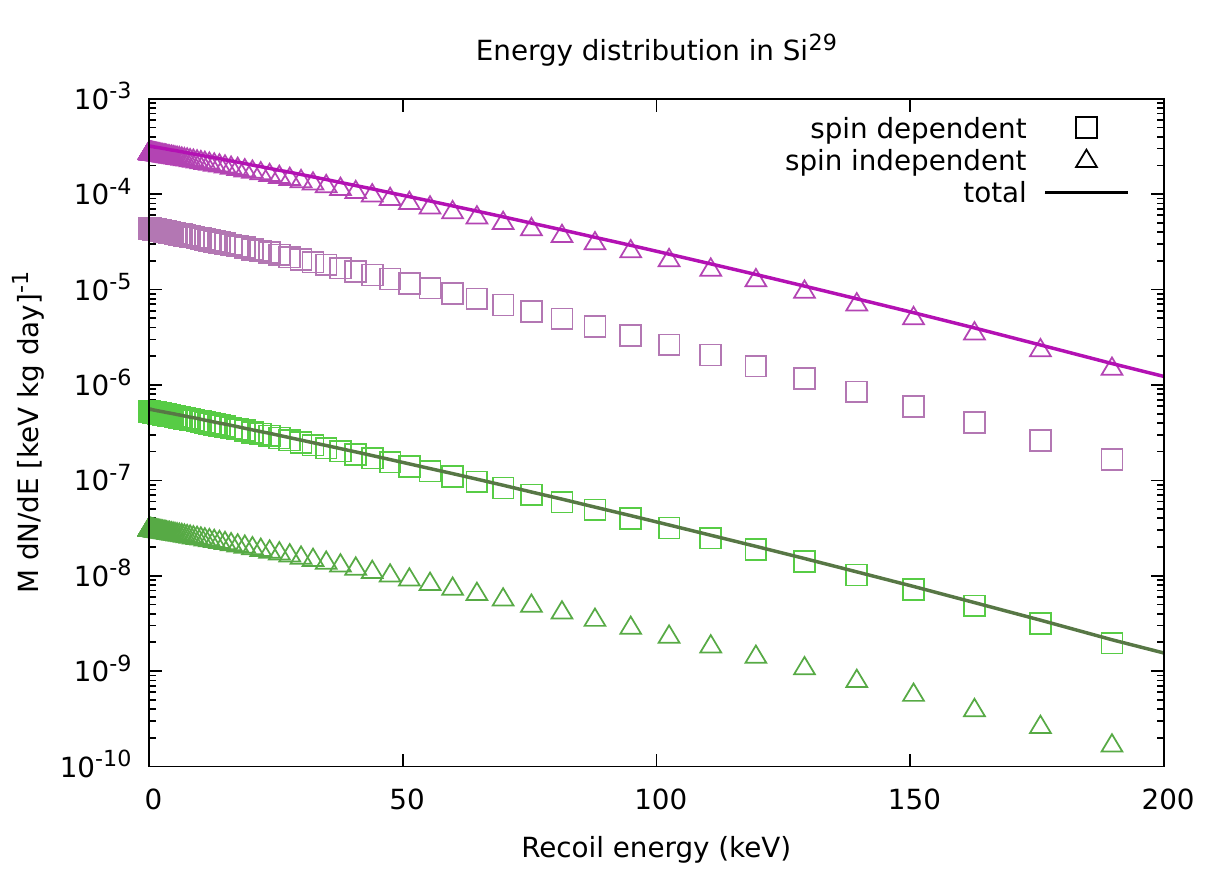}
 	\caption{Event rate (times WIMP mass) versus recoil energy for the selected BPs 66, 69 and 72 in detector material made of (from top to bottom) Xe, Ge, Na and Si, for both the active (purple) and inert (green) neutralino.}
\label{recoils}
\end{figure}
\begin{figure}
 \includegraphics[scale=0.4]{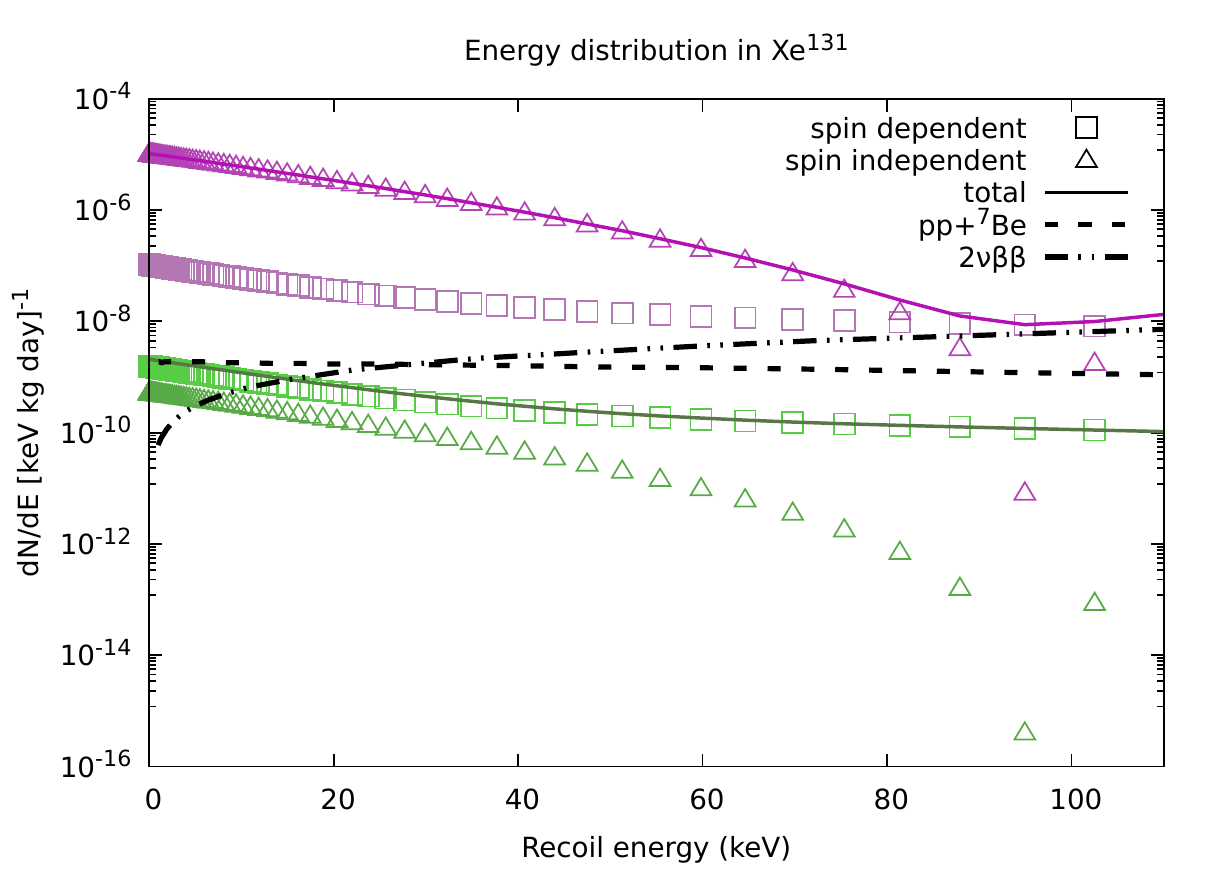}
 \includegraphics[scale=0.4]{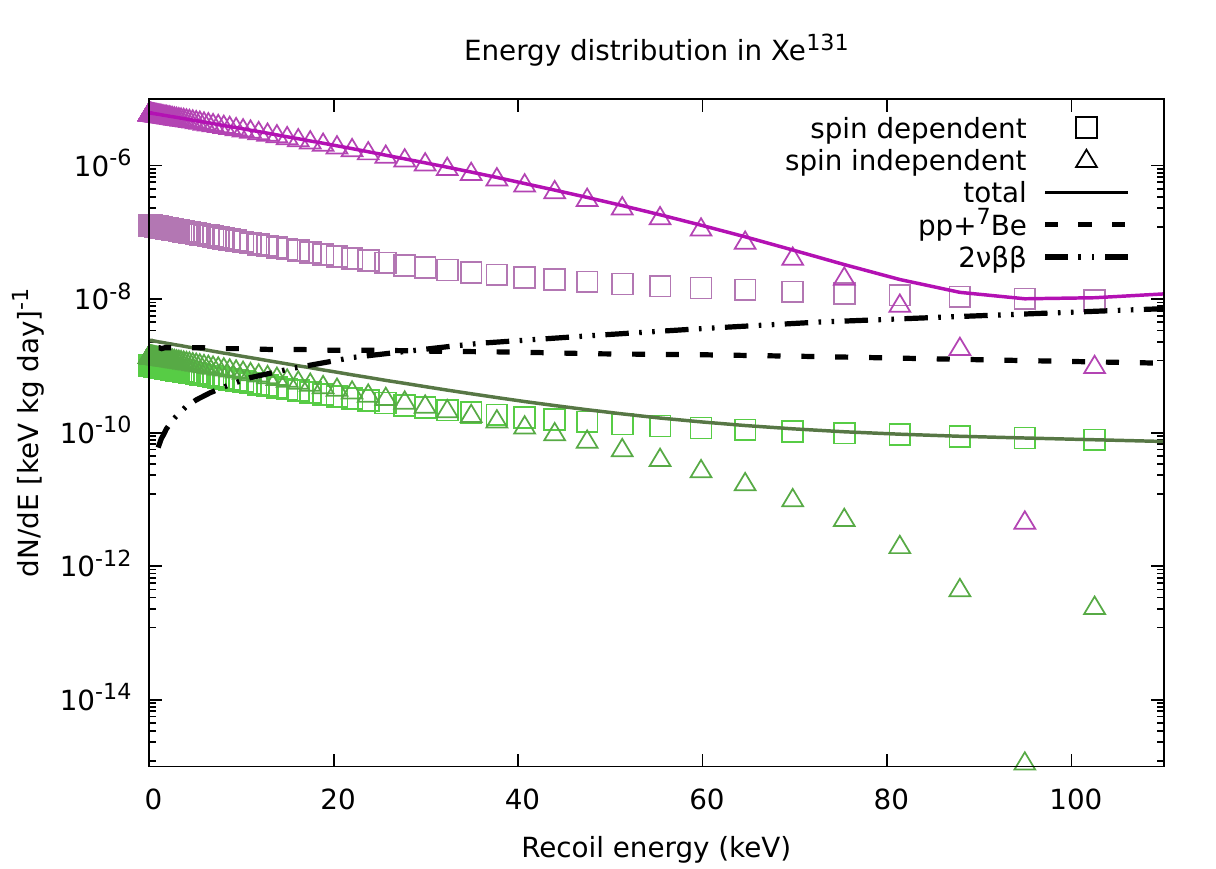}
 \includegraphics[scale=0.4]{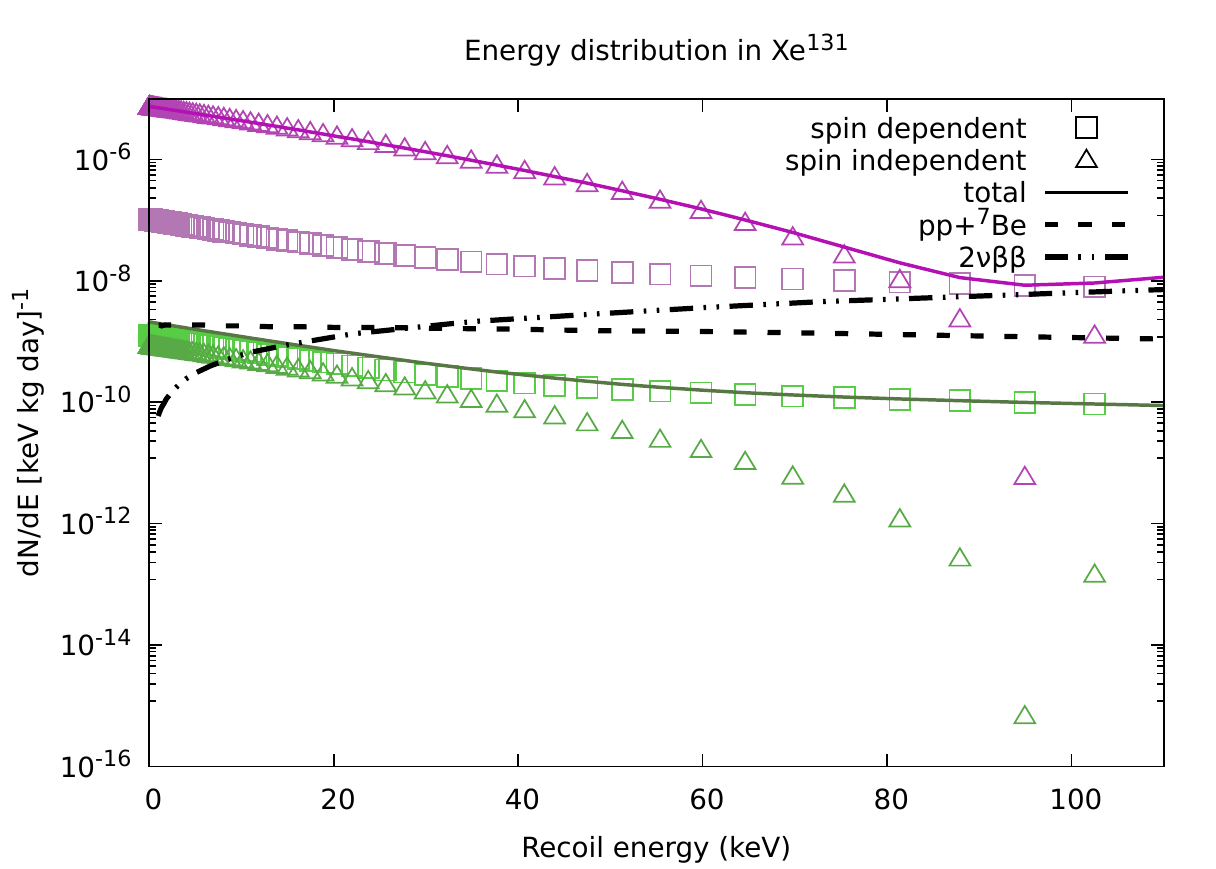}\\
 \includegraphics[scale=0.4]{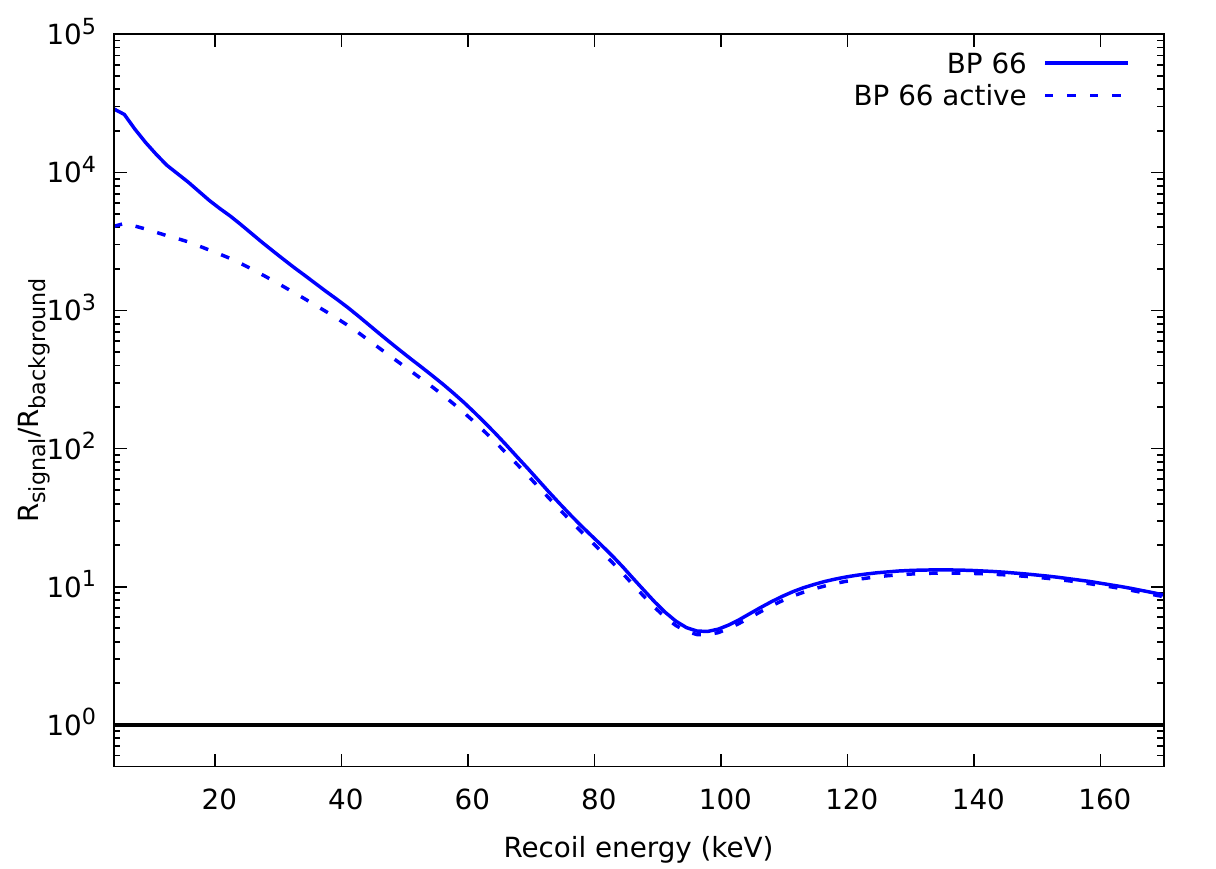}
 \includegraphics[scale=0.4]{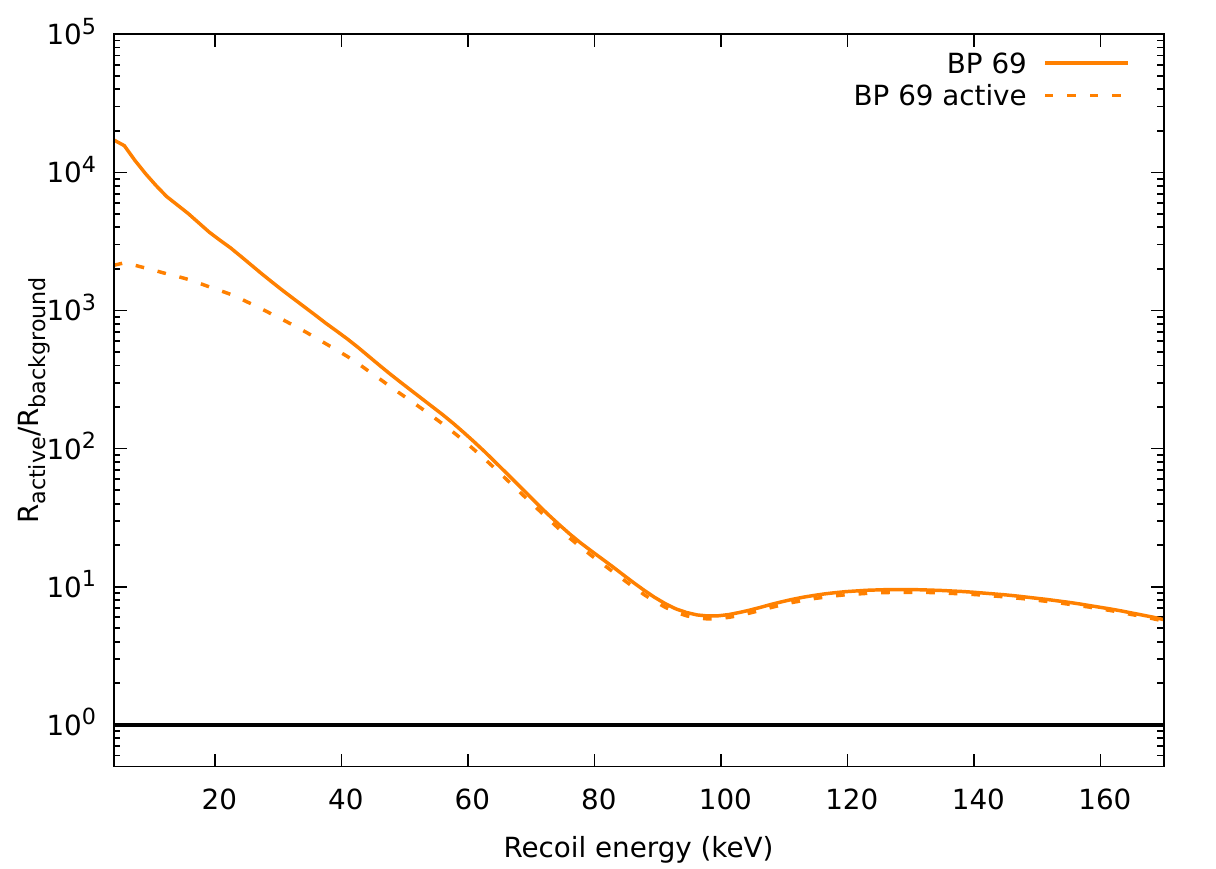}
 \includegraphics[scale=0.4]{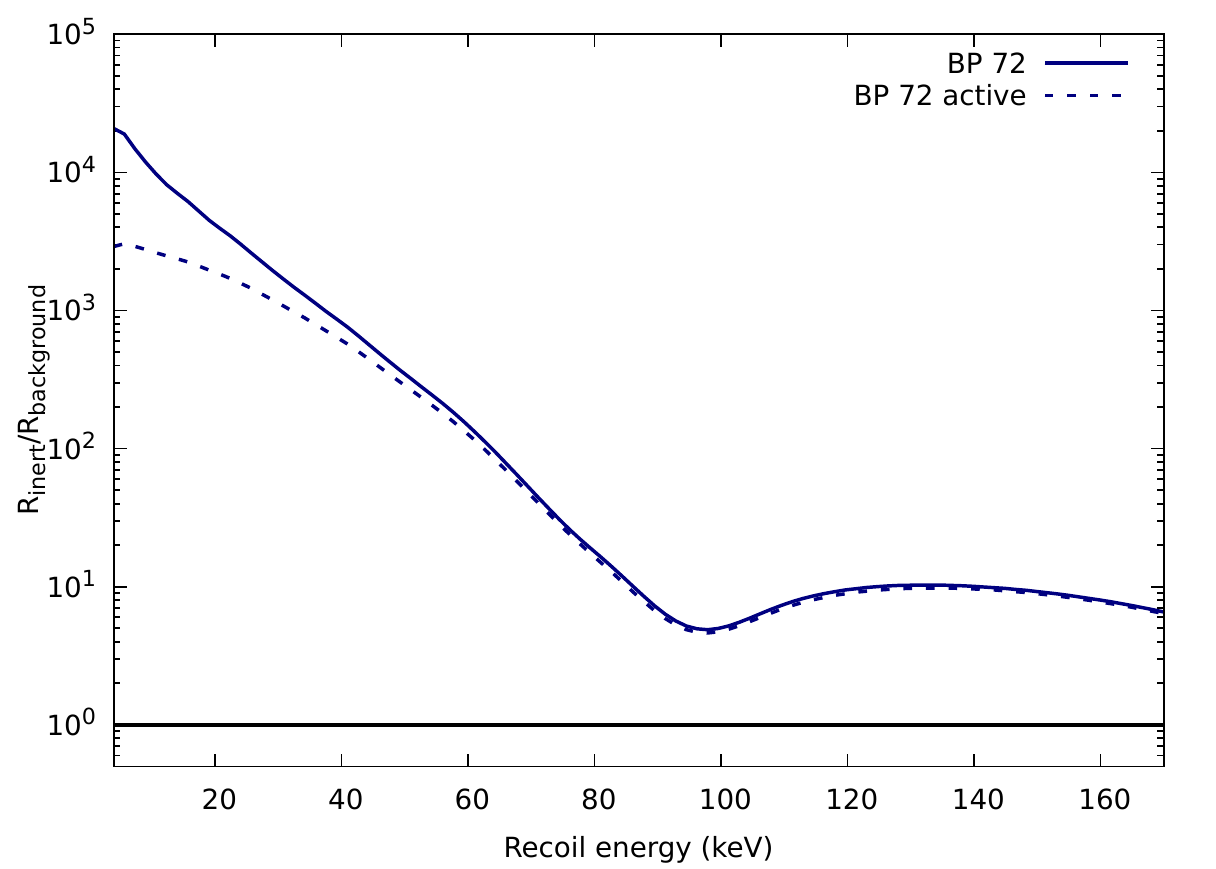}
	\caption{(Top) Nuclear recoil spectrum for the selected BPs 66, 69 and 72 in a liquid Xe detector. Also shown are the expected backgrounds \cite{Aalbers2016,Baudis2014}. (Bottom) Ratio of active plus inert neutralino (solid) and active neutralino only (dotted) signal rates to the total background ones as obtained from the top plots.}
		\label{Xebkg}
\end{figure}

In the case of SI interactions, the distribution of the number of events versus the recoil energy can be calculated as
\be 
	\frac{dN^{\text{\rm SI}}}{dE} = \frac{2M_{\text{dec}}t}{\pi}\frac{\rho_0}{M_\chi}F^2_A(q)(\lambda_p Z + \lambda_n(A - Z))^2 I(E),
\ee
where $\rho_0$ is the DM density near the Earth, $M_{\text{dec}}$ the mass of the detector, $t$ the exposure time and $F_A(q)$ the nucleus FF which depends on the momentum transfer $q=\sqrt{2EM_A}$ and
\bea
	I(E) &=& \int^\infty_{v_{\text{min}}(E)}\frac{f(v)}{v}dv, \\
	v_{\text{min}}(E) &=& \left(\frac{EM_A}{2\mu^2_\chi}\right)^{1/2}.
\eea
where $M_A$ is the mass of the nucleus.

For SI interactions, the FF is a Fourier transform of the nucleus distribution function,
\be 
	F_A(q) = \int e^{-iqx}\rho_A(x)d^3x,
\ee
where $\rho_A(x)$ is normalised such that $F_A(0)=1$. In \verb|micrOMEGAs| \cite{Belanger2009}, the Fermi distribution function is used:
\be 
	\rho_A(r) = \frac{c_{\text{norm}}}{1 + e^{(r - R_A)/a}},
\ee
where the normalisation condition fixes $c_{\text{norm}}$, and $R_A = 1.23A^{\frac{1}{3}}-0.6$ fm for a surface thickness $a=0.52$ fm.

In the case of SD interactions, instead of the nuclear FF $F_A(q)$ one needs  to introduce instead the coefficients $S_{00}(q)$, $S_{11}(q)$ and $S_{01}(q)$, which are the nuclear structure functions that take into account both the magnitude of the spin in the nucleon and the spatial distribution of it. Then, the number of events over the nucleus recoil energy is
\be 	
	\frac{dN^{\text{\rm SD}}}{dE} = \frac{8M_{\text{dec}}t}{2J_A + 1}\frac{\rho_0}{M_\chi}(S_{00}(q)a^2_0 + S_{01}(q)a_0a_1 + S_{11}(q)a^2_1) I(E),
\ee
where the SD FFs are described using a Gaussian distribution:
\be 
	S_{ij}(q) = S_{ij}(0)e^{-q^2R_A^2/4},
\ee
where $R_A=1.7A^{\frac{1}{3}}-0.28-0.78\left(A^{\frac{1}{3}}-3.8+\sqrt{(A^{\frac{1}{3}}-3.8)^2+0.2}\right)$ fm.
The normalisation of $S_{ij}(0)$ and the list of all SD FFs implemented in \verb|micrOMEGAs| can be consulted in \cite{Belanger2009}.

In Fig. \ref{recoils} we show the event rate for DD of  active higgsinos (purple) and inert higgsinos (green) as a function of the recoil energy of the nucleus for different dectection materials: Xe, Ge, Na and Si. The rate is multiplied by the WIMP mass to make the curves on the plot independent of the neutralino mass. This recoil energy can be split in its SI and SD parts. As the SI part for the inert DM candidate is much lower than the SD part (in comparison with the active DM candidate), the total shape for both DM candidates is different. The expected sensitivity for experiments using Ge, Na or Si is not low enough for our DM candidates to be seen in these cases \cite{Akerib2010}. For Xe, it is different. Thus, in the top three frames of Fig. \ref{Xebkg}, we show the expected sensitivity to see the DM candidates in BPs 66, 69 and 72 in the case of liquid Xe experiments, wherein we have a background that includes the electron recoil spectrum from the double-beta decay of $^{136}$Xe ($2\nu\beta\beta$) and the summed differential energy for $pp$ and $^7$Be neutrinos ($pp+^7$Be) undergoing neutrino-electron scattering. 
To discriminate between Nuclear Recoil (NR) and Electronic Recoil (ER) or background, liquid Xe experiments split the signal in two regions, S1 and S2, which respond differently to such recoils \cite{Cushman2013}.
In here, a 99.98\% discrimination of ERs at 30\% NR acceptance is assumed and the recoil energies are derived using the S1 signal only, according to Refs. \cite{Aalbers2016,Baudis2014}. Finally, in the three bottom frames of  Fig. \ref{Xebkg}, we show the ratio of the rates for  total signal (alongside that of the active neutralino only) and background. Specifically, the total signal rate corresponds to the addition of the SI and SD event rates of both the active and inert higgsino while the background rate corresponds to the addition of the neutrino rates from double beta decay and $\nu e$ scattering. The total number of events corresponding to our DM candidates is up to $10^{4}$ times larger that the number of events with neutrinos for low recoil energies and up to ten times larger for recoil energies larger that 100 keV, thus clearly vouching for a forthcoming (potential) discovery of the active neutralino component of DM. Furthermore, the difference in shape between the latter and the total signal for recoil energies below 60 keV may offer a hint of the presence of a second DM component, so long that a significant level of control can be achieved on the shapes of both the dominant DM signal and the background.

\section{Two Higgsinos DM and Indirect Detection Experiments}\label{sec:ID}

\begin{table}
	\begin{tabular}{|c|c|c|c|c|}
		\hline
		BP & mass (GeV) & $\langle\sigma^{\rm ann}_{WW} v\rangle(cm^3/s)$ & $\langle\sigma^{\rm ann}_{ZZ} v\rangle(cm^3/s)$ & $\langle\sigma^{\rm ann}_{h_1Z} v\rangle(cm^3/s)$ \\
		\hline
		66 (active) & 903 & $6.67\times 10^{-27}$ & $5.03\times 10^{-27}$ & $5.94\times 10^{-28}$ \\
		66 (inert) & 606 & $1.45\times 10^{-26}$ & $1.18\times 10^{-26}$ & $4.74\times 10^{-29}$ \\
		69 (active) & 619 & $1.41\times 10^{-26}$ & $1.09\times 10^{-26}$ & $1.16\times 10^{-27}$ \\  
		66 (inert) & 926 & $6.67\times 10^{-27}$ & $5.03\times 10^{-27}$ & $5.94\times 10^{-29}$ \\
		72 (active) & 766 & $9.51\times 10^{-27}$ & $7.66\times 10^{-27}$ & $7.68\times 10^{-29}$ \\
		72 (inert) & 753 & $9.48\times 10^{-27}$ & $7.64\times 10^{-27}$ & $7.62\times 10^{-29}$ \\
		\hline			 
	\end{tabular}
	\caption{The main annihilation channels of the higgsinos in our selected BPs. These are the values of $\langle\sigma^{\rm ann}_iv\rangle$ in Eq. (\ref{Eq:DFDM}).}
	\label{iDD-Xsections}
\end{table}

One of the most important methods for detecting  DM in the galactic halo is the observation of $\gamma$-rays emitted from the DM annihilation at the galactic center. The differential spectrum of  the total observed $\gamma$-ray flux is given by 
\be
\frac{d\Phi_{\rm tot}}{dE_\gamma}=\frac{d\Phi_{\gamma}}{dE_\gamma}+\frac{d\Phi_{\rm BG}}{dE_\gamma}, 
\label{Eq:DFSB}
\ee
where $d\Phi_{\gamma}/dE_\gamma$ is the differential $\gamma$-ray flux generated from the DM,  defined as
\be 
 \frac{d\Phi_\gamma(E_\gamma, \psi)}{dE_{\gamma}} = \sum_i \frac{dN^i_\gamma}{d E_\gamma} \frac{\langle \sigma^{\rm ann}_i v\rangle}{8\pi m_{\tilde{\chi}_1}^2}\frac{1}{\Delta\Omega}\int_{\Delta\Omega} d\Omega \int_{\rm los} \rho^2(r)~dl,\label{Eq:DFDM} 
 \ee
where $d N^i_\gamma/dE_\gamma$ is  the $\gamma$-ray spectrum produced per annihilation $i$ and the astrophysical factor of Eq.~(\ref{Eq:DFDM}) can be identified as
\be
\langle J \rangle_{\Delta\Omega} = \int_{\Delta\Omega} d\Omega \int_{\rm los} \rho^2(r)~dl.
\ee
For generalised Navarro-Frenk-White (NFW) halo profile with inner slope $\gamma=1.3$, one finds that the astrophysical factor $\langle J \rangle_{\Delta\Omega}$ is of order ${\cal O}(10^{22})$~GeV$^2$~cm$^{-5}$ \cite{Daylan2016,Calore2015}. Finally, $d\Phi_{\rm BG}/dE_\gamma$ is the isotropic $\gamma$-ray backgrounds \cite{Abdo2010}. In Tab. \ref{iDD-Xsections} we give the values of $\langle\sigma^{\rm ann}_iv\rangle$ for each of the active and inert higgsinos of our selected BPs.

\begin{figure}
 \includegraphics[scale=0.45]{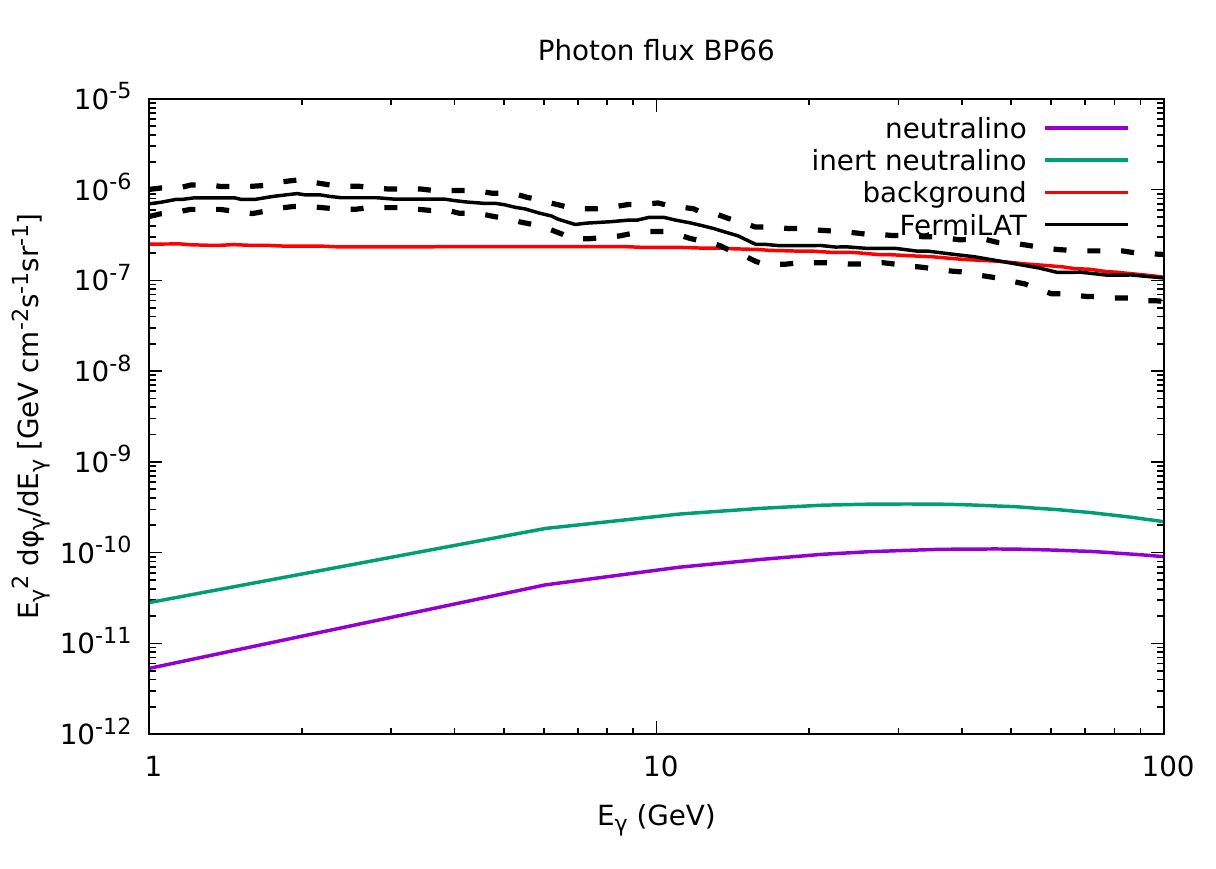}
 \includegraphics[scale=0.45]{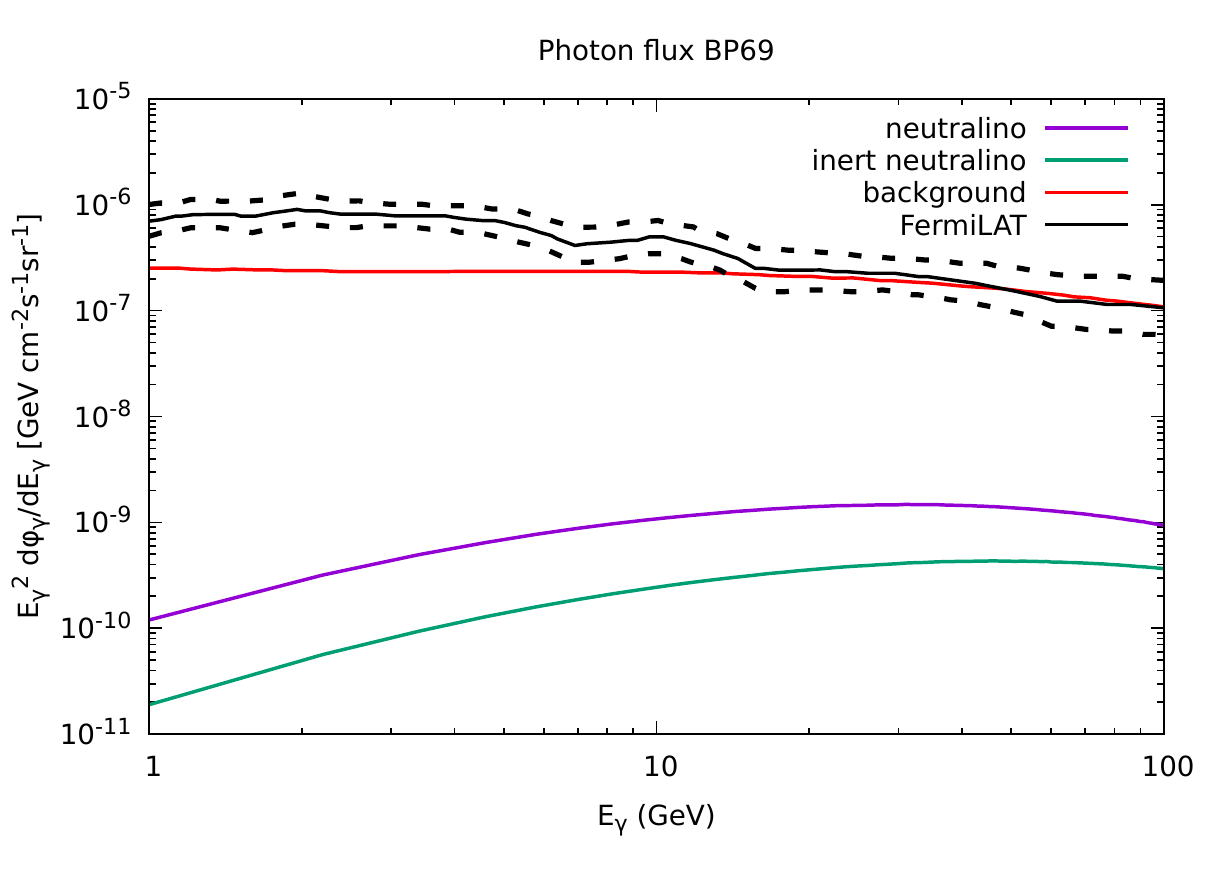}
 \includegraphics[scale=0.45]{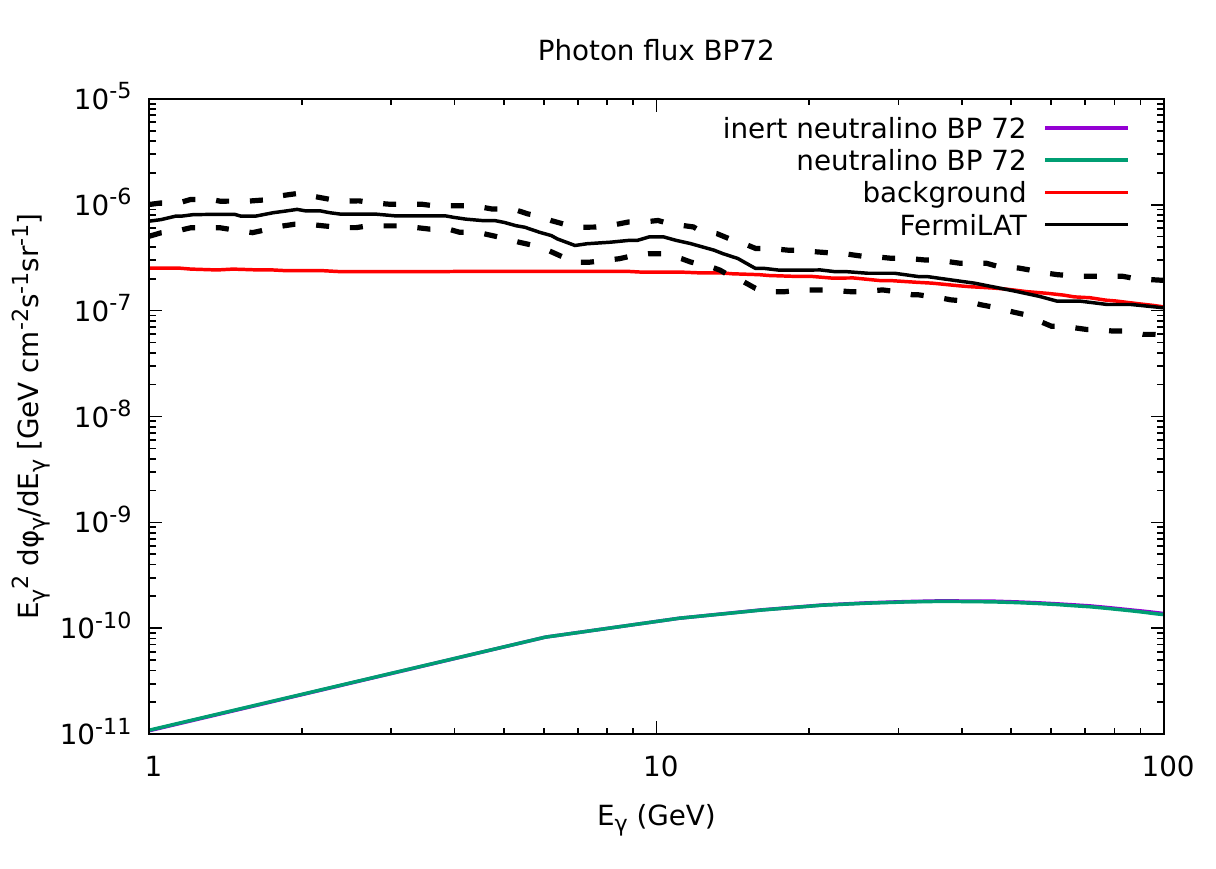}
 \caption{Differential flux of $\gamma$-ray secondary radiation from the galactic center as a function of the photon energy. We show the $\gamma$-ray spectrum produced by the three exemplary BPs 66, 69 and 72 for the active neutralino (purple) and the inert neutralino (green), respectively. The FermiLAT data (with error) is presented in black and the corresponding distribution for the background is shown in red (from Ref. \cite{DelleRose2018}). When the two DM candidates are of similar mass, as in BP 72, the two distributions are identical.}
\label{gammaflux}
\end{figure}

In Fig. \ref{gammaflux}, we show the differential flux of $\gamma$-ray secondary radiation from the galactic center as a function of the photon energy. We show the $\gamma$-ray spectrum produced by the three exemplary BPs 66, 69 and 72 for both the active  
(purple) and inert (green) neutralino. The FermiLAT data (with error) are  presented in black and the corresponding distribution for the background is shown in red. The distributions are very dependent on the masses and the flux is larger the lighter the mass of the higgsino. In contrast with DD experiments, when the inert neutralino is the lightest it would be this the DM candidate giving the largest contribution to the photon flux.

The $\gamma$-ray flux can be evaluated as
\be
	\Phi_\gamma(E,\phi) = \frac{\sigma v}{m^2_{\text{DM}}}f_\gamma(E)H(\phi)
\ee
and is expressed in number of photons per cm$^2$ per s per sr. The factor $H$ includes the integral of the squared of the DM density over the line of sight,
\be 
	H(\phi)=\frac{1}{8\pi}\int^\infty_0 dr\bar{\rho^2}(r^\prime),
\ee
where $r^\prime = \sqrt{r^2 + r^2_{\odot} - 2rr_{\odot}\cos\phi}$ and $\phi$ being the angle of observation, $r_\odot$ the distance from the Sun to the galactic center.

Another two interesting astrophysical probes to analyse with DM candidates are the antiproton and positron excesses reported by PAMELA \cite{Galper2018} and AMS-02 \cite{Aguilar2016}. The propagation of charged particles in Cosmic Rays (CRs) is one of the largest sources of uncertainties in predicting the background of antiprotons or positrons and the signal from DM annihilation \cite{Belanger2011, Aguilar2016, Cui2018, Donato2004}. Charged particles are deflected by a diffusion process in the random galactic magnetic field. The equation that describes the evolution of the energy distribution is
\be
	\frac{\partial}{\partial z}(V_C\psi_a) - \nabla\cdot(K(E)\nabla\psi_a) - \frac{\partial}{\partial E}(b(E)\psi_a) = Q_a(\mathbf{x},E),
	\label{Eq:chargedflux}
\ee 
where $\psi_a = dN/dE$ is the number density of particles per unit volume and energy, $a$ denotes the particle species, $Q_a$ is the source term and $b(E)$ the energy loss rate. For antiprotons there is an extra term which accounts for a negative contribution to the source term, as we explain below. Further, $K$ is the space diffusion coefficient, assumed homogeneous:
\be 
	K(E) = K_0\beta(E)(\mathcal{R}/1~{\rm GV})^\delta,
\ee
where $\beta$ is the particle velocity and $\mathcal{R} = p/q$ its rigidity. 
The collision between primary nuclei and the interstellar gas leads
to fragmentation of the parent nuclei and the production of secondary nuclei. Therefore the secondary-to-primary particle ratio, e.g., the Boron-to-Carbon (B/C) ratio, is usually employed to constrain the propagation parameters \cite{Maurin2001,Feng2016,Cui2018}.  The coefficient $K(E)$ introduced above  is adequate to fit this B/C data.

In the case of positrons flux, the loss rate is dominated by synchrothon radiation in the galactic magnetic field and inverse Compton scattering on stellar light and CMB photons, so that it is defined by
\be 
	b(A) = \frac{E^2}{E_0\tau_E},
\ee
where $\tau_E = 10^{16}$ s is the typical energy loss time. Then the positron flux from DM annihilation reads as 
\be 
	\psi_{\bar{e}}(E_0,r_\odot,0) = \frac{\sigma v}{b(E_0)}\int^{m_{\tilde{\chi}}}_{E_0}dEf(E)D(t(E_0) - t(E),r_\odot),
\ee
where $D(\tau,r_\odot)$ is a universal function for all energies. In \verb|MicrOMEGAs|, the routine \verb|posiFluxTab| tabulates first $D$ as a function of $\tau$ in the region $0 \leq \tau \leq t(E_{\rm min}) - t(m_{\tilde{\chi}})$ and then perfom a fast integration for all energies.

In the case of antiprotons, the propagation is dominated by diffusion and the effect of the galactic wind. As stated above, for antiprotons it is needed to add to Eq. (\ref{Eq:chargedflux}) a negative source, which corresponds to the fragmentation or decay of antiprotons in the interstellar medium $(H, He)$. The annihilation rate is
\be 	
	\Gamma_{\rm tot} = \sigma^{\rm ann}_{\bar{p}H}v_{\bar{p}}n_H +\sigma^{\rm ann}_{\bar{p}He}v_{\bar{p}}n_{He},
\ee
where $v_{\bar{p}}$ is the velocity of the antiproton, $n_H = 0.9$ cm$^{-3}$ and $n_{He} = 0.1$ cm$^{-3}$ are the average densities in the galactic disc while $\sigma^{\rm ann}_{\bar{p}H(e)}$ is the annihilation cross section that can be found in Ref.  \cite{Belanger2011,Tan1982}. (The production of the secondary antiprotons is not included in the \verb|MicrOMEGAs| code.)

Afer integration, the antiproton energy spectrum is
\be 	
	\psi_{\bar{p}}(E,r_\odot ,0) = \frac{\sigma v f_{\bar{p}}(E)}{K}\int^\infty_0 rdr\int^L_0 dz\bar{G}(r,z,0)e^{-k_cz}\int^\pi_0 d\phi\frac{\overline{\rho^2(r,z')}}{m^2_{\tilde{\chi}}},
\ee with $r' = \sqrt{r^2_\odot + r^2 + 2r_\odot r\cos\phi}$ and where $\bar{G}(r,z,0)$ is the Green function which determines the probability of a CR to propagate from the source to the detector, as defined in Ref. \cite{Belanger2011}.

\begin{figure}
 \includegraphics[scale=0.45]{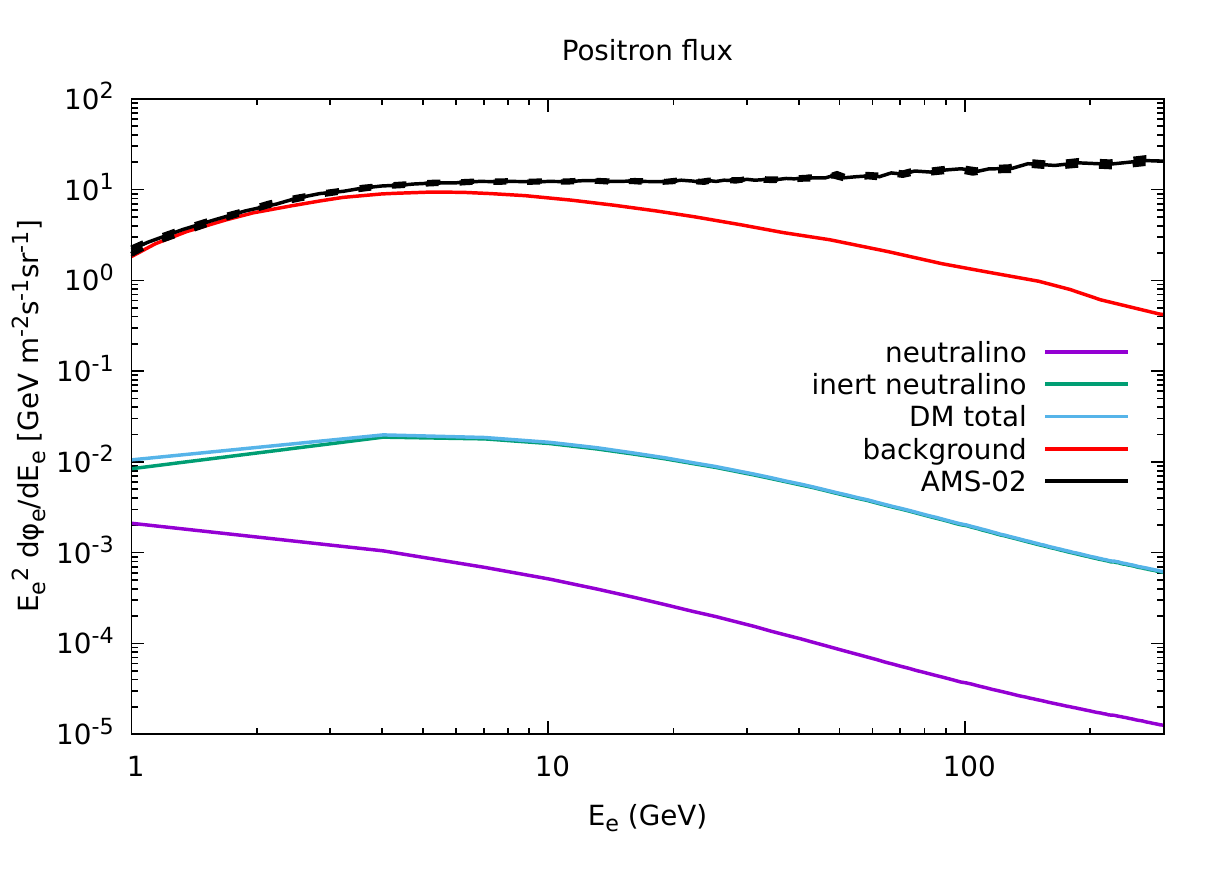}
 \includegraphics[scale=0.45]{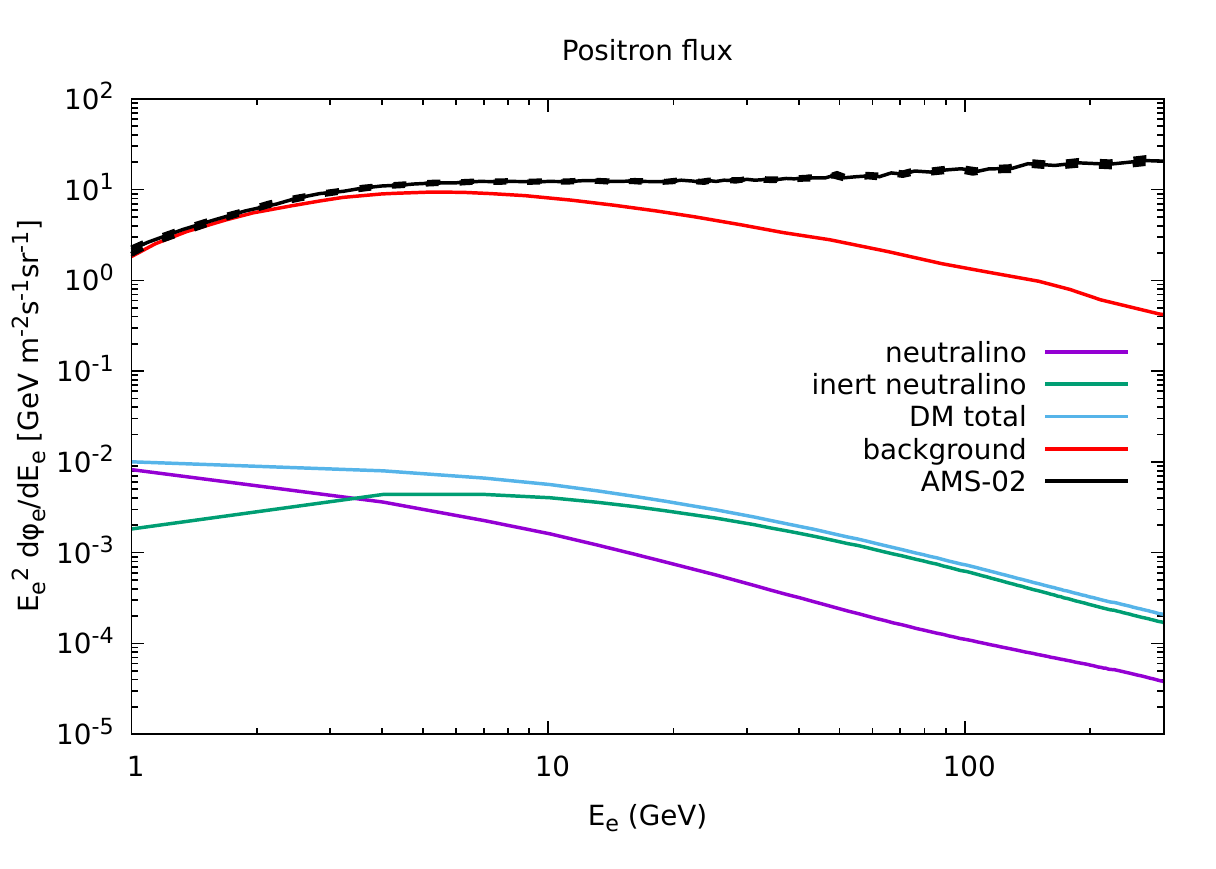}
 \includegraphics[scale=0.45]{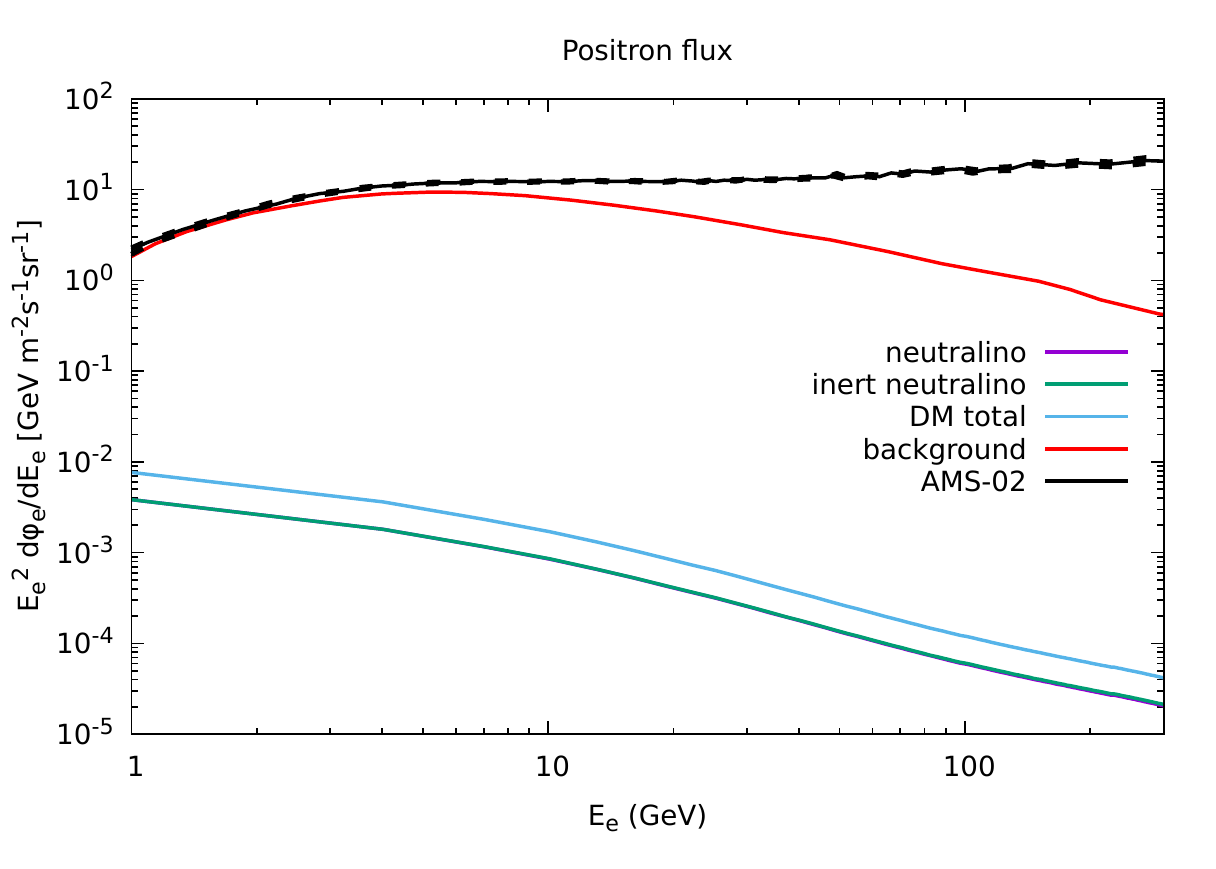}
 \caption{Positron flux versus energy produced by the three exemplary BPs 66, 69 and 72 for the active neutralino (purple) and the inert neutralino (green), also shown the addition of the two (DM total, cyan), respectively. The AMS-02 data (with error) is presented in black and the corresponding diffuse background is shown in red (from Ref. \cite{Queiroz2020}).}
\label{positronflux}
\end{figure}

\begin{figure}
 \includegraphics[scale=0.45]{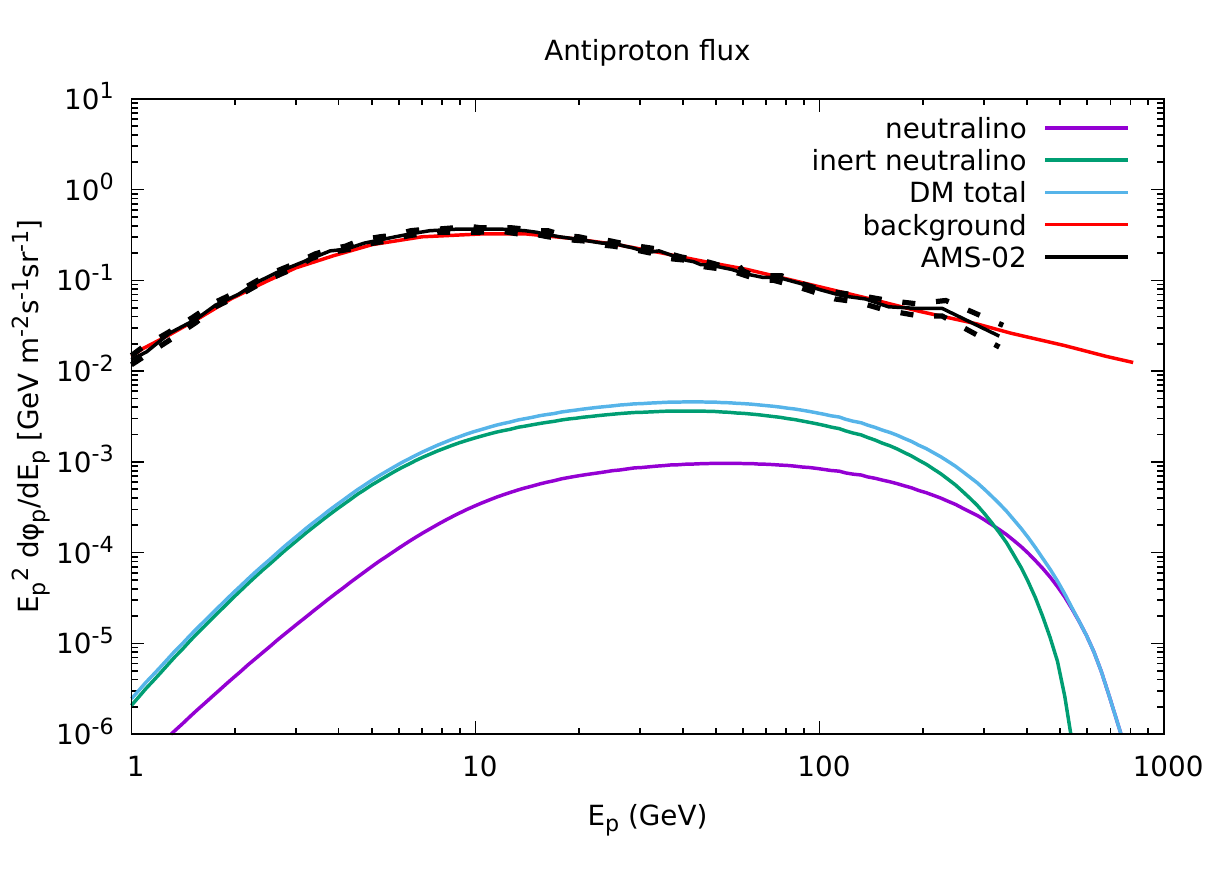}
 \includegraphics[scale=0.45]{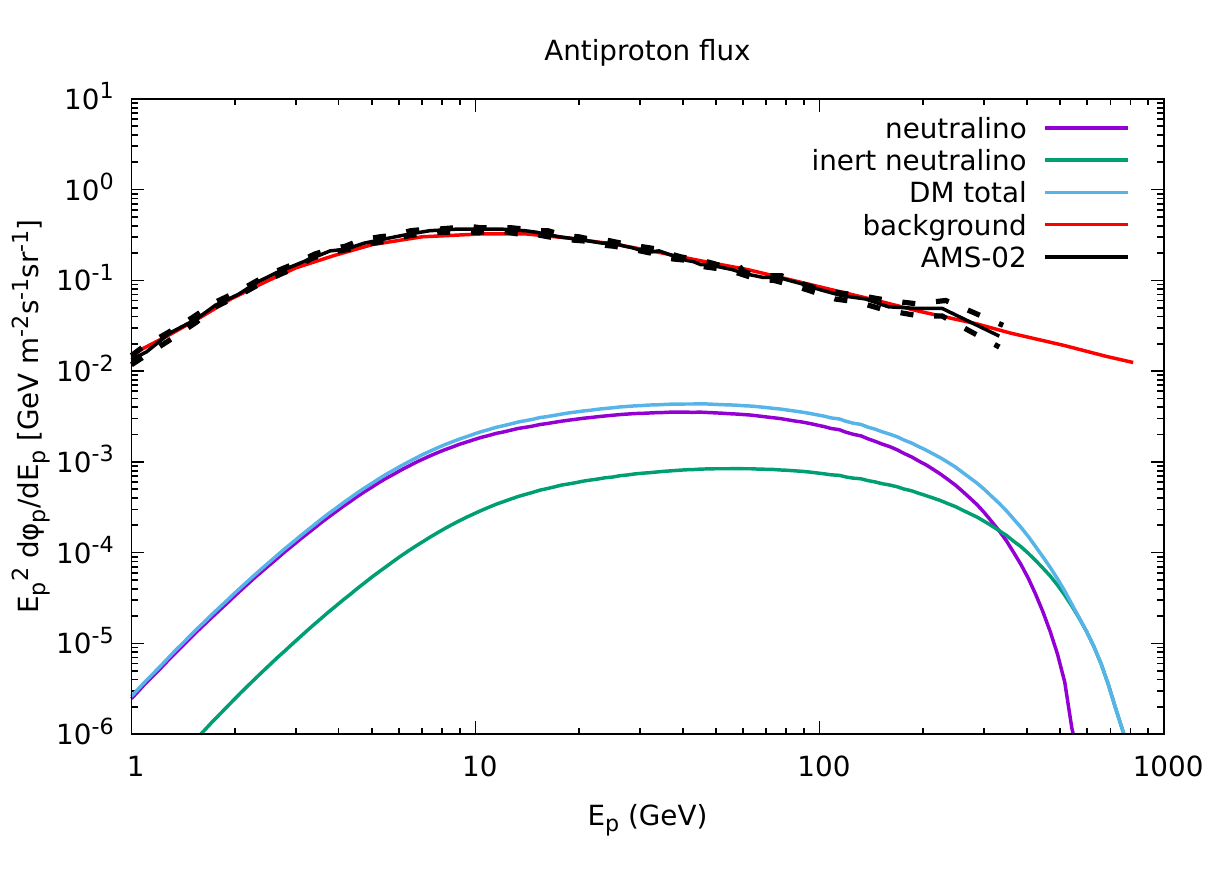}
 \includegraphics[scale=0.45]{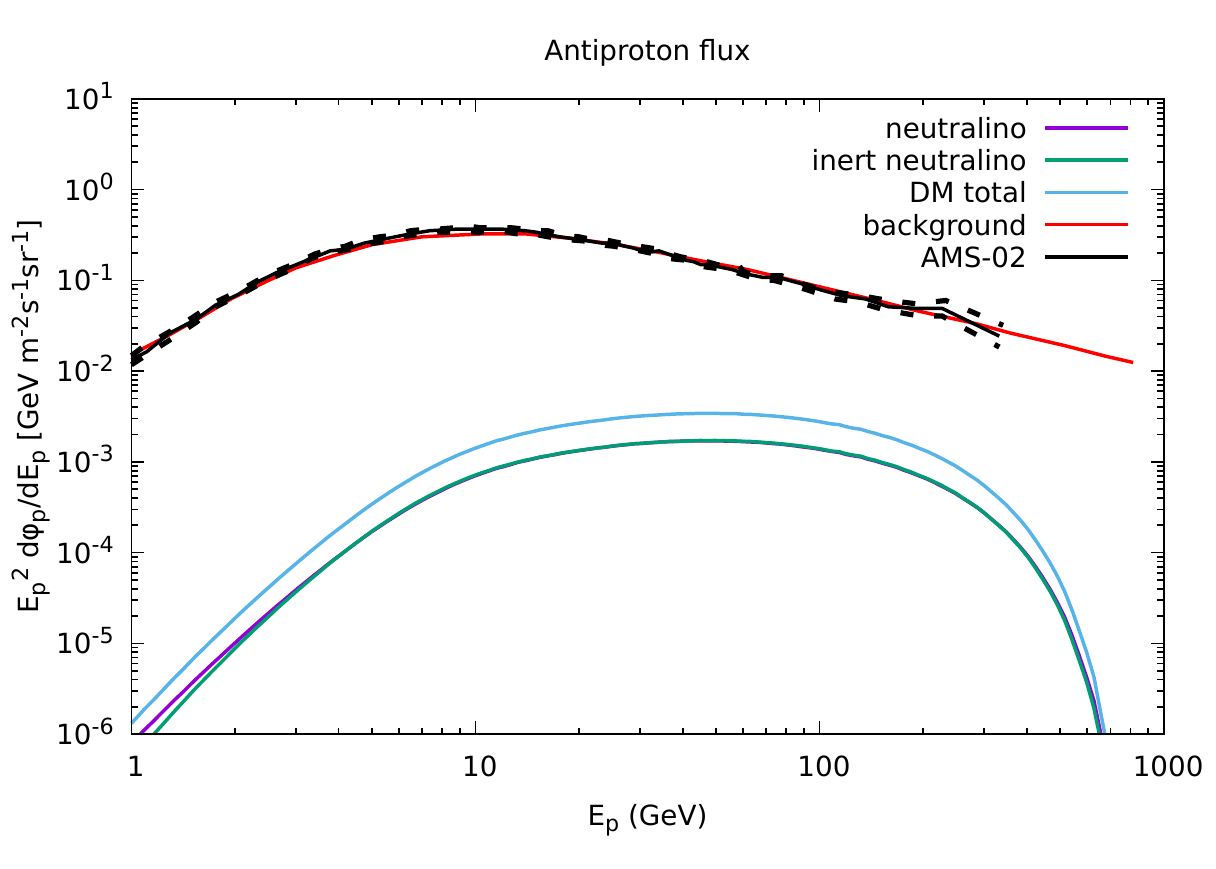}
 \caption{Antiproton flux versus energy produced by the three exemplary BPs 66, 69 and 72 for the active neutralino (purple) and the inert neutralino (green). Also shown is the addition of the two (DM total, cyan). The AMS-02 data (with errors) are presented in black and the corresponding background is shown in red (from Ref. \cite{Cui2018}).}
\label{antiprotonflux}
\end{figure}

In Figs. \ref{positronflux} and \ref{antiprotonflux} we show the resulting fluxes for  positron and antiproton production, respectively. For both cases we are comparing with the recent AMS-02 data \cite{Aguilar2019}. In the case of the positron flux, the data reported by AMS-02 seems compatible with the background (given by a diffused flux which corresponds to the observations of positrons before the last AMS-02 report \cite{Queiroz2020}) at low energies, but it requires a new source to explain the flux at higher energies. Unfortunately, both the active and inert higgsinos have a shape similar to the background and fail in proving a source to explain the excess at energies above 10 GeV. In the case of the antiproton flux, the background corresponds to inelastic collisions between protons and the interstellar medium \cite{Cui2018}. The data of AMS-02 fits  very well with the background, although there is room to improve the fit for energies between 10 to 100 GeV. However, a better explanation to this excess would correspond to a DM candidate of lighter mass than the active and inert higgsinos of the E$_6$SSM.



%
\section{Conclusions}
\label{sec:summa}

We have analysed the possibility of having two-component DM in a  simplified version of a SUSY GUT model, the E$_6$SSM, which is inspired by string theory. We have shown that one of these DM components is the lightest active neutralino (higgsino-like) that has direct couplings to the SM fermions. The other DM component is the lightest inert neutralino (also higgsino-like) that does not have a direct interaction with the SM fermions, however, it couples to the $SU(2)$ gauge bosons as it is originating from the $SU(2)$ inert Higgs doublets. These two particles are stable, hence, they are candidates for DM, because of the $R$-parity and $Z_2$ symmetries to which they obey, respectively. 

We have then emphasised that the relic abundance limit $\Omega h^2 = 0.12 \pm 0.002$ implies that the sum of the masses of these two 
DM components is nearly constant and of order $1.5$ TeV. We have thus considered three BPs  that give comparable contributions to $\Omega h^2$ from each DM component. In particular, we have studied the following cases: $m_{\chi_{\rm active}} > m_{\chi_{\rm inert}}$, $m_{\chi_{\rm active}} \approx m_{\chi_{\rm inert}}$ and $m_{\chi_{\rm active}} < m_{\chi_{\rm inert}}$. We have first investigated the DD of these three cases. In particular, we have considered the SI and SD cross sections of these two DM components and compared their results with current and future experimental measurements. We have shown that the active higgsino is within the future XENON-nT and DARWIN sensitivity, however, the inert higgsino cross sections are generally too low, which make its identification
extremely difficult, though not altogether impossible. In addition, we have studied the nuclear recoil energy associated with the two DM components while interacting detector materials: Xe, Ge, Na and Si. Here, we have shown that, again, for future Xe detectors, the active higgsino results can be well above the background (especially at low energies), hence, that this DM component has a chance to be probed in future DD experiments also for the purpose of extracting its mass. In contrast, the inert higgsino contributions are quite small and lower than the background, although this DM candidate may become evident in the recoil distribution of Xe experiments for significant exposures, when both the background and other DM signal will be known more accurately, in the form of a modifiation to the dominant signal shape (at small to intermediate recoil energies). Finally, we have analysed the ID rates, through $\gamma$-ray spectra as well as positron and antiproton fluxes, for both active and inert higgsinos. In this case, we have found that all results stemming from active DM dynamics are well below the current limits and the inert DM predictions are extremely small in comparisons, so that we have concluded that sensitivity here will be minimal for years to come.

\section*{Acknowledgements}

DR-C  is  supported  by  the  Royal  Society Newton International Fellowship NIF/R1/180813. 
SK acknowledges partial support from the Durham IPPP Visiting  Academics  (DIVA)  programme.   SM  is  financed  in  part  through  the  NExT Institute  and  the  STFC  consolidated  Grant  No.   ST/L000296/1.
HW acknowledges financial support from the Magnus Ehrnrooth Foundation and STFC Rutherford International Fellowship (funded through the MSCA-COFUND-FP Grant No.  665593). We would like to thank Yasar Hiciylmaz for helpful discussions and for providing the code to produce Figure 6. The authors acknowledge the use of the IRIDIS High Performance Computing Facility, and associated support services at the University of Southampton, in the completion of this work.


\begin{thebibliography}{100}

\bibitem{Ade2016}
{\scshape Planck} collaboration, \emph{Planck2015 results},
  \href{https://doi.org/10.1051/0004-6361/201525830}{\emph{Astronomy {\&}
  Astrophysics} {\bfseries 594} (2016) A13}
  [\href{https://arxiv.org/abs/1502.01589}{{\ttfamily 1502.01589}}].

\bibitem{Aghanim2018}
{\scshape Planck} collaboration, \emph{{Planck 2018 results. VI. Cosmological
  parameters}},  \href{https://arxiv.org/abs/1807.06209}{{\ttfamily
  1807.06209}}.

\bibitem{Wang2017}
L.~Wang, V.~Gonzalez-Perez, L.~Xie, A.P.~Cooper, C.S.~Frenk, L.~Gao et~al.,
  \emph{The galaxy population in cold and warm dark matter cosmologies},
  \href{https://doi.org/10.1093/mnras/stx788}{\emph{Monthly Notices of the
  Royal Astronomical Society} {\bfseries 468} (2017) 4579}
  [\href{https://arxiv.org/abs/1612.04540}{{\ttfamily 1612.04540}}].

\bibitem{ShaabanKhalil2019}
S.~Khalil and S.~Moretti, \emph{{Supersymmetry Beyond Minimality}}, Taylor \&
  Francis Ltd (2019).

\bibitem{King2006}
S.F.~King, S.~Moretti and R.~Nevzorov, \emph{Theory and phenomenology of an
  exceptional supersymmetric standard model},
  \href{https://doi.org/10.1103/physrevd.73.035009}{\emph{Physical Review D}
  {\bfseries 73} (2006) 035009}
  [\href{https://arxiv.org/abs/hep-ph/0510419}{{\ttfamily hep-ph/0510419}}].

\bibitem{King2006a}
S.~King, S.~Moretti and R.~Nevzorov, \emph{Exceptional supersymmetric standard
  model}, \href{https://doi.org/10.1016/j.physletb.2005.12.070}{\emph{Physics
  Letters B} {\bfseries 634} (2006) 278}
  [\href{https://arxiv.org/abs/hep-ph/0511256}{{\ttfamily hep-ph/0511256}}].

\bibitem{King2020}
S.F.~King, S.~Moretti and R.~Nevzorov, \emph{{A Review of the Exceptional
  Supersymmetric Standard Model}},
  \href{https://doi.org/10.3390/sym12040557}{\emph{Symmetry} {\bfseries 12}
  (2020) 557} [\href{https://arxiv.org/abs/2002.02788}{{\ttfamily
  2002.02788}}].

\bibitem{Athron2009}
P.~Athron, S.F.~King, D.J.~Miller, S.~Moretti and R.~Nevzorov, \emph{{{The
  Constrained Exceptional Supersymmetric Standard Model}}},
  \href{https://doi.org/10.1103/physrevd.80.035009}{\emph{Phys. Rev. D}
  {\bfseries 80} (2009) 035009}
  [\href{https://arxiv.org/abs/0904.2169}{{\ttfamily 0904.2169}}].

\bibitem{Athron2009a}
P.~Athron, S.~King, D.~Miller, S.~Moretti and R.~Nevzorov, \emph{{Predictions
  of the Constrained Exceptional Supersymmetric Standard Model}},
  \href{https://doi.org/10.1016/j.physletb.2009.10.051}{\emph{Physics Letters
  B} {\bfseries 681} (2009) 448}
  [\href{https://arxiv.org/abs/0901.1192}{{\ttfamily 0901.1192}}].

\bibitem{Athron2011}
P.~Athron, S.F.~King, D.J.~Miller, S.~Moretti and R.~Nevzorov, \emph{{LHC}
  signatures of the constrained exceptional supersymmetric standard model},
  \href{https://doi.org/10.1103/physrevd.84.055006}{\emph{Physical Review D}
  {\bfseries 84} (2011) 055006}
  [\href{https://arxiv.org/abs/1102.4363}{{\ttfamily 1102.4363}}].

\bibitem{Athron2012}
P.~Athron, S.F.~King, D.J.~Miller, S.~Moretti and R.~Nevzorov,
  \emph{Constrained exceptional supersymmetric standard model with a higgs
  signal near 125~{GeV}},
  \href{https://doi.org/10.1103/physrevd.86.095003}{\emph{Physical Review D}
  {\bfseries 86} (2012) 095003}
  [\href{https://arxiv.org/abs/1206.5028}{{\ttfamily 1206.5028}}].

\bibitem{Athron2016}
P.~Athron, D.~Harries, R.~Nevzorov and A.G.~Williams, \emph{{$E_6$ Inspired
  SUSY benchmarks, dark matter relic density and a 125 GeV Higgs}},
  \href{https://doi.org/10.1016/j.physletb.2016.06.040}{\emph{Physics Letters
  B} {\bfseries 760} (2016) 19}
  [\href{https://arxiv.org/abs/1512.07040}{{\ttfamily 1512.07040}}].

\bibitem{Athron2017}
P.~Athron, A.~Thomas, S.~Underwood and M.~White, \emph{{Dark matter candidates
  in the constrained Exceptional Supersymmetric Standard Model}},
  \href{https://doi.org/10.1103/physrevd.95.035023}{\emph{Physical Review D}
  {\bfseries 95} (2017) 035023}
  [\href{https://arxiv.org/abs/1611.05966}{{\ttfamily 1611.05966}}].

\bibitem{Hall2011}
J.P.~Hall and S.F.~King, \emph{{Bino Dark Matter and Big Bang Nucleosynthesis
  in the Constrained $E_6$SSM with Massless Inert Singlinos}},
  \href{https://doi.org/10.1007/JHEP06(2011)006}{\emph{JHEP} {\bfseries 06}
  (2011) 006} [\href{https://arxiv.org/abs/1104.2259}{{\ttfamily 1104.2259}}].

\bibitem{Hall2009}
J.P.~Hall and S.F.~King, \emph{Neutralino dark matter with inert higgsinos and
  singlinos},
  \href{https://doi.org/10.1088/1126-6708/2009/08/088}{\emph{Journal of High
  Energy Physics} {\bfseries 2009} (2009) 088}
  [\href{https://arxiv.org/abs/0905.2696}{{\ttfamily 0905.2696}}].

\bibitem{Aguila1988}
F.~del Aguila, G.~Coughlan and M.~Quiros, \emph{{Gauge Coupling Renormalization
  With Several U(1) Factors}},
  \href{https://doi.org/10.1016/0550-3213(88)90266-0}{\emph{Nucl. Phys. B}
  {\bfseries 307} (1988) 633}.

\bibitem{Aguila1988a}
F.~del Aguila, J.~Gonzalez and M.~Quiros, \emph{{Renormalization Group Analysis
  of Extended Electroweak Models From the Heterotic String}},
  \href{https://doi.org/10.1016/0550-3213(88)90265-9}{\emph{Nucl. Phys. B}
  {\bfseries 307} (1988) 571}.

\bibitem{Accomando2013}
E.~Accomando, D.~Becciolini, A.~Belyaev, S.~Moretti and
  C.~Shepherd-Themistocleous, \emph{Z $\prime$ at the {LHC}: interference and
  finite width effects in drell-yan},
  \href{https://doi.org/10.1007/jhep10(2013)153}{\emph{Journal of High Energy
  Physics} {\bfseries 2013} (2013) 153}
  [\href{https://arxiv.org/abs/1304.6700}{{\ttfamily 1304.6700}}].

\bibitem{Accomando2020}
E.~Accomando, F.~Coradeschi, T.~Cridge, J.~Fiaschi, F.~Hautmann, S.~Moretti
  et~al., \emph{Production of z$\prime$-boson resonances with large width at
  the {LHC}},
  \href{https://doi.org/10.1016/j.physletb.2020.135293}{\emph{Physics Letters
  B} {\bfseries 803} (2020) 135293}
  [\href{https://arxiv.org/abs/1910.13759}{{\ttfamily 1910.13759}}].

\bibitem{Aad2019}
{\scshape ATLAS} collaboration, \emph{Search for high-mass dilepton resonances
  using 139 fb$^{-1}$ of $pp$ collision data collected at $\sqrt{s}=$13 tev
  with the atlas detector},
  \href{https://doi.org/10.1016/j.physletb.2019.07.016}{\emph{Phys. Lett. B}
  {\bfseries 796} (2019) 68}
  [\href{https://arxiv.org/abs/1903.06248}{{\ttfamily 1903.06248}}].

\bibitem{Frank2020}
M.~Frank, Y.~Hi{\c{c}}y{\i}lmaz, S.~Moretti and O.~Ozdal, \emph{E$_6$ motivated
  {UMSSM} confronts experimental data},
  \href{https://doi.org/10.1007/jhep05(2020)123}{\emph{Journal of High Energy
  Physics} {\bfseries 2020} (2020) 123}
  [\href{https://arxiv.org/abs/2004.01415}{{\ttfamily 2004.01415}}].

\bibitem{Profumo2004}
S.~Profumo and C.E.~Yaguna, \emph{Statistical analysis of supersymmetric dark
  matter in the minimal supersymmetric standard model after {WMAP}},
  \href{https://doi.org/10.1103/physrevd.70.095004}{\emph{Physical Review D}
  {\bfseries 70} (2004) 095004}
  [\href{https://arxiv.org/abs/hep-ph/0407036}{{\ttfamily hep-ph/0407036}}].

\bibitem{Randall1999}
L.~Randall and R.~Sundrum, \emph{Out of this world supersymmetry breaking},
  \href{https://doi.org/10.1016/s0550-3213(99)00359-4}{\emph{Nuclear Physics B}
  {\bfseries 557} (1999) 79}
  [\href{https://arxiv.org/abs/hep-th/9810155}{{\ttfamily hep-th/9810155}}].

\bibitem{Bhattacharya2013}
S.~Bhattacharya, A.~Drozd, B.~Grzadkowski and J.~Wudka, \emph{{Two-Component
  Dark Matter}}, \href{https://doi.org/10.1007/JHEP10(2013)158}{\emph{JHEP}
  {\bfseries 10} (2013) 158} [\href{https://arxiv.org/abs/1309.2986}{{\ttfamily
  1309.2986}}].

\bibitem{Staub2014}
F.~Staub, \emph{{SARAH 4 : A tool for (not only SUSY) model builders}},
  \href{https://doi.org/10.1016/j.cpc.2014.02.018}{\emph{Comput. Phys. Commun.}
  {\bfseries 185} (2014) 1773}
  [\href{https://arxiv.org/abs/1309.7223}{{\ttfamily 1309.7223}}].

\bibitem{Porod2012}
W.~Porod and F.~Staub, \emph{{SPheno} 3.1: extensions including flavour,
  {CP}-phases and models beyond the {MSSM}},
  \href{https://doi.org/10.1016/j.cpc.2012.05.021}{\emph{Computer Physics
  Communications} {\bfseries 183} (2012) 2458}
  [\href{https://arxiv.org/abs/1104.1573}{{\ttfamily 1104.1573}}].

\bibitem{Belanger2010}
G.~Belanger, F.~Boudjema, A.~Pukhov and A.~Semenov, \emph{{micrOMEGAs: A Tool
  for dark matter studies}},
  \href{https://doi.org/10.1393/ncc/i2010-10591-3}{\emph{Il Nuovo Cimento C}
  {\bfseries 33} (2010) 111} [\href{https://arxiv.org/abs/1005.4133}{{\ttfamily
  1005.4133}}].

\bibitem{Belanger2015}
G.~B{\'{e}}langer, F.~Boudjema, A.~Pukhov and A.~Semenov,
  \emph{{micrOMEGAs}4.1: Two dark matter candidates},
  \href{https://doi.org/10.1016/j.cpc.2015.03.003}{\emph{Computer Physics
  Communications} {\bfseries 192} (2015) 322}
  [\href{https://arxiv.org/abs/1407.6129}{{\ttfamily 1407.6129}}].

\bibitem{Aprile2018}
{\scshape XENON} collaboration, \emph{{Dark Matter Search Results from a One
  Ton-Year Exposure of XENON1T}},
  \href{https://doi.org/10.1103/PhysRevLett.121.111302}{\emph{Phys. Rev. Lett.}
  {\bfseries 121} (2018) 111302}
  [\href{https://arxiv.org/abs/1805.12562}{{\ttfamily 1805.12562}}].

\bibitem{Jungman1996}
G.~Jungman, M.~Kamionkowski and K.~Griest, \emph{Supersymmetric dark matter},
  \href{https://doi.org/10.1016/0370-1573(95)00058-5}{\emph{Physics Reports}
  {\bfseries 267} (1996) 195}
  [\href{https://arxiv.org/abs/hep-ph/9506380}{{\ttfamily hep-ph/9506380}}].

\bibitem{Belanger2009}
G.~B{\'{e}}langer, F.~Boudjema, A.~Pukhov and A.~Semenov, \emph{Dark matter
  direct detection rate in a generic model with {micrOMEGAs}{\_}2.2},
  \href{https://doi.org/10.1016/j.cpc.2008.11.019}{\emph{Computer Physics
  Communications} {\bfseries 180} (2009) 747}
  [\href{https://arxiv.org/abs/0803.2360}{{\ttfamily 0803.2360}}].

\bibitem{Aalbers2016}
{\scshape DARWIN} collaboration, \emph{{DARWIN: towards the ultimate dark
  matter detector}},
  \href{https://doi.org/10.1088/1475-7516/2016/11/017}{\emph{JCAP} {\bfseries
  11} (2016) 017} [\href{https://arxiv.org/abs/1606.07001}{{\ttfamily
  1606.07001}}].

\bibitem{Baudis2014}
L.~Baudis, A.~Ferella, A.~Kish, A.~Manalaysay, T.~Marrodan~Undagoitia and
  M.~Schumann, \emph{{Neutrino physics with multi-ton scale liquid xenon
  detectors}}, \href{https://doi.org/10.1088/1475-7516/2014/01/044}{\emph{JCAP}
  {\bfseries 01} (2014) 044} [\href{https://arxiv.org/abs/1309.7024}{{\ttfamily
  1309.7024}}].

\bibitem{Akerib2010}
{\scshape CDMS} collaboration, \emph{{A Low-Threshold Analysis of CDMS
  Shallow-Site Data}},
  \href{https://doi.org/10.1103/PhysRevD.82.122004}{\emph{Phys. Rev. D}
  {\bfseries 82} (2010) 122004}
  [\href{https://arxiv.org/abs/1010.4290}{{\ttfamily 1010.4290}}].

\bibitem{Cushman2013}
P.~Cushman et~al., \emph{{Working Group Report: WIMP Dark Matter Direct
  Detection}},  in \emph{{Community Summer Study 2013}: {Snowmass on the
  Mississippi}}, 10, 2013 [\href{https://arxiv.org/abs/1310.8327}{{\ttfamily
  1310.8327}}].

\bibitem{Daylan2016}
T.~Daylan, D.P.~Finkbeiner, D.~Hooper, T.~Linden, S.K.N.~Portillo, N.L.~Rodd
  et~al., \emph{{The characterization of the gamma-ray signal from the central
  Milky Way: A case for annihilating dark matter}},
  \href{https://doi.org/10.1016/j.dark.2015.12.005}{\emph{Phys. Dark Univ.}
  {\bfseries 12} (2016) 1} [\href{https://arxiv.org/abs/1402.6703}{{\ttfamily
  1402.6703}}].

\bibitem{Calore2015}
F.~Calore, I.~Cholis and C.~Weniger, \emph{{Background Model Systematics for
  the Fermi GeV Excess}},
  \href{https://doi.org/10.1088/1475-7516/2015/03/038}{\emph{JCAP} {\bfseries
  03} (2015) 038} [\href{https://arxiv.org/abs/1409.0042}{{\ttfamily
  1409.0042}}].

\bibitem{Abdo2010}
{\scshape Fermi-LAT} collaboration, \emph{{The Spectrum of the Isotropic
  Diffuse Gamma-Ray Emission Derived From First-Year Fermi Large Area Telescope
  Data}}, \href{https://doi.org/10.1103/PhysRevLett.104.101101}{\emph{Phys.
  Rev. Lett.} {\bfseries 104} (2010) 101101}
  [\href{https://arxiv.org/abs/1002.3603}{{\ttfamily 1002.3603}}].

\bibitem{Galper2018}
{\scshape PAMELA} collaboration, \emph{{Ten years of CR physics with PAMELA}},
  \href{https://doi.org/10.1016/j.asr.2017.08.026}{\emph{Adv. Space Res.}
  {\bfseries 62} (2018) 2892}.

\bibitem{Aguilar2016}
{\scshape AMS} collaboration, \emph{{Antiproton Flux, Antiproton-to-Proton Flux
  Ratio, and Properties of Elementary Particle Fluxes in Primary Cosmic Rays
  Measured with the Alpha Magnetic Spectrometer on the International Space
  Station}}, \href{https://doi.org/10.1103/PhysRevLett.117.091103}{\emph{Phys.
  Rev. Lett.} {\bfseries 117} (2016) 091103}.

\bibitem{Belanger2011}
G.~Belanger, F.~Boudjema, P.~Brun, A.~Pukhov, S.~Rosier-Lees, P.~Salati et~al.,
  \emph{{Indirect search for dark matter with micrOMEGAs2.4}},
  \href{https://doi.org/10.1016/j.cpc.2010.11.033}{\emph{Comput. Phys. Commun.}
  {\bfseries 182} (2011) 842}
  [\href{https://arxiv.org/abs/1004.1092}{{\ttfamily 1004.1092}}].

\bibitem{Cui2018}
M.-Y.~Cui, X.~Pan, Q.~Yuan, Y.-Z.~Fan and H.-S.~Zong, \emph{Revisit of cosmic
  ray antiprotons from dark matter annihilation with updated constraints on the
  background model from {AMS}-02 and collider data},
  \href{https://doi.org/10.1088/1475-7516/2018/06/024}{\emph{Journal of
  Cosmology and Astroparticle Physics} {\bfseries 2018} (2018) 024}
  [\href{https://arxiv.org/abs/1803.02163}{{\ttfamily 1803.02163}}].

\bibitem{Donato2004}
F.~Donato, N.~Fornengo, D.~Maurin and P.~Salati, \emph{{Antiprotons in cosmic
  rays from neutralino annihilation}},
  \href{https://doi.org/10.1103/PhysRevD.69.063501}{\emph{Phys. Rev. D}
  {\bfseries 69} (2004) 063501}
  [\href{https://arxiv.org/abs/astro-ph/0306207}{{\ttfamily
  astro-ph/0306207}}].

\bibitem{Maurin2001}
D.~Maurin, F.~Donato, R.~Taillet and P.~Salati, \emph{{Cosmic rays below z=30
  in a diffusion model: new constraints on propagation parameters}},
  \href{https://doi.org/10.1086/321496}{\emph{Astrophys. J.} {\bfseries 555}
  (2001) 585} [\href{https://arxiv.org/abs/astro-ph/0101231}{{\ttfamily
  astro-ph/0101231}}].

\bibitem{Feng2016}
J.~Feng, N.~Tomassetti and A.~Oliva, \emph{{Bayesian analysis of
  spatial-dependent cosmic-ray propagation: astrophysical background of
  antiprotons and positrons}},
  \href{https://doi.org/10.1103/PhysRevD.94.123007}{\emph{Phys. Rev. D}
  {\bfseries 94} (2016) 123007}
  [\href{https://arxiv.org/abs/1610.06182}{{\ttfamily 1610.06182}}].

\bibitem{Tan1982}
L.~Tan and L.~Ng, \emph{{Parametrization of anti-p invariant cross-section in
  pp-collisions using a new scaling variable}},
  \href{https://doi.org/10.1103/PhysRevD.26.1179}{\emph{Phys. Rev. D}
  {\bfseries 26} (1982) 1179}.

\bibitem{Aguilar2019}
{\scshape AMS} collaboration, \emph{{Towards Understanding the Origin of
  Cosmic-Ray Positrons}},
  \href{https://doi.org/10.1103/PhysRevLett.122.041102}{\emph{Phys. Rev. Lett.}
  {\bfseries 122} (2019) 041102}.

\bibitem{Queiroz2020}
F.S.~Queiroz and C.~Siqueira, \emph{{Explaining the AMS positron excess via
  Right-handed Neutrinos}},
  \href{https://doi.org/10.1103/PhysRevD.101.075007}{\emph{Phys. Rev. D}
  {\bfseries 101} (2020) 075007}
  [\href{https://arxiv.org/abs/1910.04782}{{\ttfamily 1910.04782}}].

\bibitem{DelleRose2018}
L.D.~Rose, S.~Khalil, S.J.D.~King, S.~Kulkarni, C.~Marzo, S.~Moretti et~al.,
  \emph{{Sneutrino Dark Matter in the BLSSM}},
  \href{https://doi.org/10.1007/jhep07(2018)100}{\emph{Journal of High Energy
  Physics} {\bfseries 2018} (2018) 100}
  [\href{https://arxiv.org/abs/1712.05232}{{\ttfamily 1712.05232}}].

\end{thebibliography}

\end{document}